\documentclass[12pt]{iopart}

\usepackage[dvipdfmx]{graphicx}
\usepackage{amsmath}
\usepackage{amssymb}
\usepackage{amscd}
\usepackage{bm}

\usepackage[mathscr]{eucal}

\usepackage{graphicx, xcolor}

\newtheorem{theorem}{Theorem}
\newtheorem{lemma}[theorem]{Lemma}

\newtheorem{proposition}[theorem]{Proposition}

\newtheorem{remark}[theorem]{Remark}

\newcommand{\Z}{{\mathbb Z}}
\newcommand{\R}{{\mathbb R}}

\newcommand{\proof}{\noindent \textit{Proof. }}
\newcommand{\qed}{\hfill $\Box$}

\newcommand\uu{\mbox{\Large \textsf{\textbf{u}}}}

\newcommand{\batten}[4]{%
\begin{picture}(40,40)(-20,-20)
	\put(-10,0){\line(1,0){20}}
	\thicklines
	\put(0,10){\line(0,-1){20}}
	\put(-11,0){\makebox(0,0)[r]{$#1$}}
	\put(0,11){\makebox(0,0)[b]{$#2$}}
	\put(0,-11){\makebox(0,0)[t]{$#3$}}
	\put(11,0){\makebox(0,0)[l]{$#4$}}
\end{picture}
}

\newcommand{\battenshift}[4]{%
\begin{picture}(40,40)(-20,-20)
	\thicklines
  \qbezier[200](-10,0)(0,0)(0,10)
  \qbezier[200](0,-10)(0,0)(10,0)
	\put(-11,0){\makebox(0,0)[r]{$#1$}}
	\put(0,11){\makebox(0,0)[b]{$#2$}}
	\put(0,-11){\makebox(0,0)[t]{$#3$}}
	\put(11,0){\makebox(0,0)[l]{$#4$}}
\end{picture}
}

\begin{document}
\title{Differential equations for the closed geometric crystal chains}

\author{Taichiro Takagi}
\address{Department of Applied Physics, National Defense Academy, Yokosuka, Kanagawa 239-8686, Japan}
\ead{takagi@nda.ac.jp}
\begin{abstract}
We present two types of systems of differential equations
that can be derived from a set of discrete integrable systems
which we call the closed geometric crystal chains.
One is a kind of extended Lotka-Volterra systems, and
the other seems to be generally new but reduces to a previously known system
in a special case.
Both equations have Lax representations
associated with what are known as the loop elementary symmetric functions, which were originally
introduced to describe
products of 
affine type $A$ geometric crystals for symmetric tensor representations.
Examples of the derivations of the continuous time Lax equations from
a discrete time one are described in detail,
where a novel method of taking a continuum limit by
assuming asymptotic behaviors of the eigenvalues
of the Lax matrix in Puiseux series expansions is used. 
\end{abstract}
\title[Differential equations for the closed geometric crystal chains]
\maketitle

\section{Introduction}\label{sec:1}
{In modern mathematical physics, studies on 
the relations between continuous, discrete, and
ultra-discrete integrable dynamical systems have been developed extensively \cite{FOY00, FvML06, GKT04,IKT12,Suris03, TTMS96}.
Motivated by many studies in this field,
and in particular by those 
based on the theory of crystals \cite{Ka1,Ka2} and that of geometric crystals
\cite{BK00},} the author and T.~Yoshikawa \cite{TY21} constructed
a new class of discrete integrable systems that can be viewed as
a geometric lifting of a class of integrable cellular automata known as
the periodic box-ball systems \cite{KS08, KTT, KT10, YYT, YT02}.
Since they are related to a realization of
type $A_{n-1}$ geometric crystals
and geometric $R$-matrices by G.~Frieden \cite{F19, F21},
we called the new integrable systems {\em closed geometric crystal chains}.

From a conventional viewpoint,
the rank of the affine Lie algebra $A_{n-1}$ is related to the number of solitons 
in the periodic box-ball system.
In this viewpoint, the geometric lifting of the system has already known and
identified with the discrete periodic Toda chain \cite{HT95, KT02}.
On the other hand,  
this rank is related to the number of
degrees of freedom in each site variables in our construction.
Therefore, a closed geometric crystal chain is 
not equivalent
to the discrete periodic Toda chain.
As a result, while 
the discrete Toda chain yields the Toda equation in a continuum limit,
it is unclear what kind of differential equations will be derived
from the closed geometric crystal chains in such a limit,
%
except for the case of $n=2$
that was
studied in \S 2.3.2 of reference \cite{TY21}.

The purpose of this paper is to give an outlook for
extending this result  to the case of general $n$.
We present two types of systems of differential equations
that can be derived from the chains.
One is equation \eqref{eq:1},
which is a kind of extended Lotka-Volterra systems \cite{B88, Itoh1987, Narita1982}. 
The other is equation \eqref{eq:2}
with a function $e^{(\alpha)}_{L-1}$ in
\eqref{eq:2022jan4_2},
which seems to be generally new but reduces to a previously 
known system studied in \cite{Suris97} 
in a special case \eqref{eq:2022_mar4_1}.
An important point here is that both equations are related to Lax equations
associated with 
{the \emph{loop elementary symmetric functions}, which were originally
introduced in reference \cite{LP11} for a description of
products of 
affine type $A$ geometric crystals for symmetric tensor representations. 
(We shall use its variation in reference \cite{ILP17}. )}
Examples of the derivations of the continuous time Lax equations from
a discrete time one are described in detail,
where a novel method of taking a continuum limit by
assuming asymptotic behaviors of the eigenvalues
of the Lax matrix in Puiseux series expansions is used. %
{In those examples, the method of imposing periodic boundary conditions
on the discrete integrable systems in \cite{TY21},
which uses Perron-Frobenius theorem, 
turns out to be related to an unexpected way of taking
the continuum limit. 
We expect that it will provide a new technique for studying
the above mentioned relations between various types of integrable dynamical systems.}
In this paper we restrict ourselves to the case of
$n=4$ for such explicit derivations.
Discussions for more general cases 
will be explored in a separate publication.

The remaining part of this paper is organized as follows.
In \S \ref{sec:2} we study a system of differential equations
that can be viewed as a kind of the extended Lotka-Volterra systems.
We introduce the loop elementary symmetric functions in \S \ref{subsec:2_1},
and define a matrix $\mathcal{L}$ with elements given by them.
We present Theorem \ref{th:main}, which is the first result of this paper,
showing that the system of differential equations
leads to a Lax equation satisfied by $\mathcal{L}$ with a companion matrix $\mathcal{Y}$.
A proof of this theorem is shown in \S \ref{subsec:2_2},
where the Lax equation is decomposed into a set of Lax triads \cite{Suris04}.
In \S \ref{subsec:2_3}, we study several properties that can be derived from the Lax equation.
In \S \ref{sec:3} we introduce another system of differential equations
related to the loop elementary symmetric functions.
The second result of this paper, Theorem \ref{th:main2}, 
is presented in \S \ref{subsec:3_1} to show
that the system of differential equations
leads to a Lax equation satisfied by $\mathcal{L}$ with another 
companion matrix $\mathcal{Z}$.
A proof of this theorem is shown in \S \ref{subsec:3_2},
where the Lax equation is decomposed into another set of Lax triads.
In \S \ref{subsec:3_3}, we study several properties that can be derived from the Lax equation.
In \S \ref{sec:4}, we present a connection of the systems of differential equations
and the closed geometric crystal chains.
We give a brief review of the latter system
in \S \ref{subsec:4_1} mainly for the case that can be described by $4 \times 4$ matrices
as an example.
In \S \ref{subsec:4_2},  \S \ref{subsec:4_3}, and \S \ref{subsec:4_4},
we give discussions for deriving the two types of differential equations
from the closed geometric crystal chains in the case of $n=4$,
where Theorems \ref{th:main3} and \ref{th:main4} are presented
as the third and fourth results of this paper.
Several concluding remarks are given in \S \ref{sec:5} and
a few detailed calculations are shown in \ref{app:A} and \ref{app:B}.

\section{Type I differential equations}\label{sec:2}
\subsection{Definitions and the first result}\label{subsec:2_1}
Let $L,n$ be a pair of coprime integers.
Then there is a unique integer $0<p<n$ such that the condition 
$Lp \equiv 1 \, (\mod n)$ is satisfied.
Let $t \in \R$ be the time variable
and $u_i^{(\alpha)}$ be a set of
dependent variables 
labeled by $(\alpha, i) \in (\Z/n\Z)\times (\Z/L\Z)$.
Suppose that the system of differential equations
\begin{equation}\label{eq:1}
\frac{{\rm d} u_i^{(\alpha)}}{{\rm d} t} = u_i^{(\alpha)}
\left(
\sum_{j=1}^{\min(Lp-1, L(n-p))} (u_{i-j}^{(\alpha +j)} - u_{i+j}^{(\alpha -j)})
\right),
\end{equation}
is satisfied by them.

For any $(\alpha, i) \in (\Z/n\Z)\times (\Z/L\Z)$,
there is a unique integer $0 \leq m < n$ such that the condition
$1-i-\alpha \equiv L m \, (\mod n)$ is satisfied.
Set $f(\alpha, i) := i + Lm$, which can be viewed as an element of $\Z/(nL)\Z$.
By the Chinese remainder theorem, the map sending $(\alpha, i)$
to the $f(\alpha, i)$ gives
an isomorphism of rings
$(\Z/n\Z)\times (\Z/L\Z) \simeq \Z/(nL)\Z$, for which the inverse map is
given by sending $f$ to $((1-f) (\mod n), f (\mod L))$.
Thus if we write $P_{f(\alpha, i)} = u_i^{(\alpha)}$,
then equation \eqref{eq:1} is written as $\dot{P}_f = P_f \left(
\sum_{g=1}^{\min(Lp-1, L(n-p))} 
(P_{f-g} - P_{f+g}) \right)$, which is a kind of the extended Lotka-Volterra systems.

For the set of variables $u_i^{(\alpha)}$ and an integer $m$, 
let $e^{(\alpha)}_m \, (\alpha \in \Z/n\Z)$ be the $m$-th
{\em loop elementary symmetric functions}
defined by
\begin{equation}\label{eq:lesf}
e^{(\alpha)}_m = \sum_{1 \leq j_1 <j_2 < \dots < j_m \leq L}
u_{j_1}^{(\alpha+1-j_1)} u_{j_2}^{(\alpha+2-j_2)} \cdots u_{j_m}^{(\alpha+m-j_m)},
\end{equation}
and $e^{(\alpha)}_0 =1$, $e^{(\alpha)}_m =0 \, (m<0 \quad
\mbox{or} \quad m>L)$ \cite{ILP17,LP11}.
In particular, we have
$e^{(\alpha)}_1 =u_1^{(\alpha)}+u_2^{(\alpha -1)}+\dots+u_L^{(\alpha -L+1)}$, 
$e^{(\alpha)}_L = u_1^{(\alpha)} u_2^{(\alpha)} \dots u_L^{(\alpha)}$, and
\begin{equation}\label{eq:2022jan4_2}
e^{(\alpha)}_{L-1} = \sum_{i=1}^L \left(
\prod_{j=1}^{i-1}  u_j^{(\alpha)} \prod_{k=i+1}^{L}  u_k^{(\alpha -1)} \right).
\end{equation}
Let $\lambda$ be an indeterminate and
$\mathcal{L}$ be the $n \times n$ matrix
defined by
\begin{equation}\label{eq:Lmatrix}
\mathcal{L} =(\mathcal{L}_{ij})_{1 \leq i,j \leq n}, \qquad
\mathcal{L}_{ij}=\sum_{m \geq 0} e^{(i)}_{j-i+L-mn}  \lambda^m,
\end{equation}
which we call a {\em Lax matrix}.
We define
\begin{equation}\label{eq:2022apr14_1}
y^{(\alpha)} = \sum_{j=0}^{p-1} e^{(\alpha -jL)}_1 - (p/n) \sum_{r =1}^{n} e^{(r)}_1,
\end{equation}
and let $\mathcal{Y}$ be the $n \times n$ matrix
\begin{equation}\label{eq:2021dec28_2}
\mathcal{Y}=
\begin{pmatrix}
y^{(1)} & & & & \lambda \\
1 & y^{(2)} & & & \\
& 1 & \ddots & & \\
&& \ddots & \ddots &\\
 & & & 1 & y^{(n)} 
\end{pmatrix}.
\end{equation}
Through the variables $u_i^{(\alpha)}$, the elements of these matrices are
functions of the time variable $t$.

\begin{theorem}\label{th:main}
Suppose that the variables $u_i^{(\alpha)}$ are satisfying the system of
differential equations \eqref{eq:1}.
Then the Lax matrix satisfies the equation
\begin{equation}\label{eq:2021dec28_3}
\frac{{\rm d} \mathcal{L}}{{\rm d} t} = 
[\mathcal{L}, \mathcal{Y}].
\end{equation}
\end{theorem}

This result implies that conserved quantities of the dynamical system
represented by equation \eqref{eq:1} can be
given by the coefficients of the characteristic polynomial of the
Lax matrix $\mathcal{L}$, or equivalently by 
$\mathrm{Tr}\, \mathcal{L}^m/m \, (m=1,\dots,n)$.

\subsection{Proof of Theorem \ref{th:main}}\label{subsec:2_2}
We define
\begin{equation}\label{eq:2022jan4_1}
\uu_i = (u_i^{(1)}, \dots, u_i^{(n)}), \quad U = (\uu_1,\uu_2,  \dots, \uu_L),
\end{equation}
and $\sigma U = (\uu_2,  \dots, \uu_L, \uu_1)$, where $\sigma$ denotes the cyclic
shift to the left.
For the $y^{(\alpha)}$ defined in \eqref{eq:2022apr14_1}, we write its dependence on
the variable $U$ as $y^{(\alpha)}(U)$, and define
\begin{equation}\label{eq:2022apr14_2}
y^{(\alpha)}_i = y^{(\alpha)}(\sigma^{i-1} U),
\end{equation}
for any $i \in \Z/L\Z$.
Let $\mathcal{Y}_i, \mathcal{M}_i$ denote the
$n \times n$ matrices defined by
\begin{equation}\label{eq:matricesBM}
\mathcal{Y}_i=
\begin{pmatrix}
y^{(1)}_i & & & & \lambda \\
1 & y^{(2)}_i & & & \\
& 1 & \ddots & & \\
&& \ddots & \ddots &\\
 & & & 1 & y^{(n)}_i 
\end{pmatrix},
\qquad
\mathcal{M}_i=
\begin{pmatrix}
u^{(1)}_i & & & & \lambda \\
1 & u^{(2)}_i & & & \\
& 1 & \ddots & & \\
&& \ddots & \ddots &\\
 & & & 1 & u^{(n)}_i 
\end{pmatrix}.
\end{equation}
Using the identities $\mathcal{L} = \mathcal{M}_1 \cdots \mathcal{M}_L$ 
(Lemma 6.~1 of \cite{ILP17}) and
$\mathcal{Y} = \mathcal{Y}_{1}$, one sees that
the assertion of Theorem \ref{th:main} follows from:
\begin{proposition}\label{prop:eqMY}
The system of differential equations \eqref{eq:1} is equivalent to
\begin{equation}\label{eq:Lax_triads}
\frac{{\rm d} \mathcal{M}_i}{{\rm d} t} = \mathcal{M}_i\mathcal{Y}_{i+1}-
\mathcal{Y}_{i}\mathcal{M}_i .
\end{equation}
\end{proposition}

\proof
The matrix elements of equation \eqref{eq:Lax_triads} are explicitly written as
\begin{align}
y^{(\alpha)}_{i+1} - y^{(\alpha +1)}_i &= u_i^{(\alpha)}  - u_i^{(\alpha +1)}, \label{eq:2021dec28_7yy}\\
\frac{{\rm d} u_i^{(\alpha)}}{{\rm d} t}&= u_i^{(\alpha)} (y^{(\alpha)}_{i+1} - y^{(\alpha)}_i).
\label{eq:2021dec28_7}
\end{align}
Thus the assertion of the proposition is a consequence of the following two lemmas,
which are satisfied by any set of variables $u_i^{(\alpha)}$ for $(\alpha, i) \in (\Z/n\Z)\times (\Z/L\Z)$.
\qed

\begin{lemma}
Let $y^{(\alpha)}_{i}$ be the one defined in \eqref{eq:2022apr14_2}.
Then the relation \eqref{eq:2021dec28_7yy} holds.
\end{lemma}

\proof
Using the expression $e^{(\alpha)}_1 =\sum_{k=0}^{L-1} u_{1+k}^{(\alpha-k)}$, we can write
$e^{(\alpha -jL)}_1 =\sum_{k=0}^{L-1} u_{1+k}^{(\alpha-jL-k)} = 
\sum_{k=0}^{L-1} u_{1+k+jL}^{(\alpha-jL-k)}$.
Therefore
\begin{equation}\label{eq:2021dec28_1}
y^{(\alpha)} (U)= \sum_{j=0}^{p-1} \sum_{k=0}^{L-1} u_{1+k+jL}^{(\alpha-jL-k)} - p c_0=
\sum_{j=0}^{Lp-1} u_{1+j}^{(\alpha-j)} - p c_0,
\end{equation}
where $c_0 := (1/n) \sum_{r =1}^{n} e^{(r)}_1$.
Hence we have $y^{(\alpha)}_{i+1} - y^{(\alpha +1)}_i =y^{(\alpha)} (\sigma^i U)-
y^{(\alpha +1)} (\sigma^{i-1} U) = \sum_{j=0}^{Lp-1} \left( u_{1+i+j}^{(\alpha-j)} - 
u_{i+j}^{(\alpha+1-j)} \right) = u_{i+Lp}^{(\alpha -Lp + 1)}  - u_i^{(\alpha +1)}= u_i^{(\alpha)}  - u_i^{(\alpha +1)}$, where the condition
$Lp \equiv 1 \, (\mod n)$ is used in the last step.
\qed

\begin{lemma}
Let $y^{(\alpha)}_{i}$ be the one defined in \eqref{eq:2022apr14_2}.
Then the following relation holds:
\begin{equation}\label{eq:2022feb8_3}
y^{(\alpha)}_{i+1} - y^{(\alpha)}_i = \sum_{j=1}^{\min(Lp-1, L(n-p))} \left( u_{i-j}^{(\alpha+j)} - 
u_{i+j}^{(\alpha-j)} \right).
\end{equation}
\end{lemma}
\proof
First we suppose $p \leq \frac{n}{2}$.
Then $\min(Lp-1, L(n-p)) = Lp-1$.
Using \eqref{eq:2021dec28_1} and the relation $u_{i+Lp}^{(\alpha -Lp + 1)} = u_i^{(\alpha)}$, we have
\begin{equation}\label{eq:2021dec28_6}
y^{(\alpha)}_{i+1} - y^{(\alpha)}_i = y^{(\alpha)} (\sigma^i U)-
y^{(\alpha)} (\sigma^{i-1} U) = \sum_{j=0}^{Lp-1} \left( u_{1+i+j}^{(\alpha-j)} - 
u_{i+j}^{(\alpha-j)} \right)=
\sum_{j=1}^{Lp-1} \left( u_{i+j}^{(\alpha-j+1)} - 
u_{i+j}^{(\alpha-j)} \right).
\end{equation}
Replacing the index $j$ by $Lp-j$ and using the condition $Lp \equiv 1 \, (\mod n)$, we have
\begin{math}
\sum_{j=1}^{Lp-1}  u_{i+j}^{(\alpha-j+1)} =
\sum_{j=1}^{Lp-1}  u_{i+Lp-j}^{(\alpha-Lp+j+1)} =\sum_{j=1}^{Lp-1}  u_{i-j}^{(\alpha+j)},
\end{math}
so that \eqref{eq:2022feb8_3} follows.

Second we suppose $p > \frac{n}{2}$.
Then $\min(Lp-1, L(n-p)) = L(n-p)$.
Since $L$ and $n$ are coprime, we have $\sum_{j \in (\Z/n\Z)} e^{(\alpha -jL)}_1 =
\sum_{\alpha =1}^{n} e^{(\alpha)}_1 = n c_0$.
This implies that
\begin{equation}\label{eq:2022mar8_1}
y^{(\alpha)} (U)= \sum_{j=0}^{p-1} e^{(\alpha -jL)}_1 - p c_0=
(n-p)c_0 - \sum_{j=1}^{n-p} e^{(\alpha +jL)}_1.
\end{equation}
Using the expression $e^{(\alpha+L)}_1 =\sum_{k=0}^{L-1} u_{1+k}^{(\alpha+L-k)}=\sum_{k=1}^{L} u_{1-k}^{(\alpha+k)}$, one can rewrite
the summation in the second term of the right hand side of 
equation \eqref{eq:2022mar8_1}
as
\begin{equation*}
\sum_{j=1}^{n-p} e^{(\alpha +jL)}_1 =
\sum_{j=1}^{L(n-p)} u_{1-j}^{(\alpha+j)} =
\sum_{j=1}^{L(n-p)} u_{j}^{(\alpha-j)},
\end{equation*}
where the last expression is derived from the second one by replacing $j$ by
$L(n-p)+1-j$.
Hence we have
$y^{(\alpha)} (\sigma^{i-1} U) =(n-p)c_0 -  \sum_{j=1}^{L(n-p)} u_{i-j}^{(\alpha+j)}$ and
$y^{(\alpha)} (\sigma^{i} U) =(n-p)c_0 -  \sum_{j=1}^{L(n-p)} u_{i+j}^{(\alpha-j)}$,
so that \eqref{eq:2022feb8_3} follows.
\qed
\begin{remark}
Let the loop elementary symmetric functions $e^{(\alpha)}_i$ be denoted by
$e^{(\alpha)}_i(U)$ for showing their dependence on the variable $U$.
Suppose $p=1$ or the condition $L \equiv 1 \, (\mod n)$ is satisfied.
Then the set of differential equations \eqref{eq:1} is written as
\begin{equation}\label{eq:2022feb8_1}
\frac{{\rm d} u_i^{(\alpha)}}{{\rm d} t} = u_i^{(\alpha)}
\left(
e^{(\alpha)}_1(\sigma^i U) - e^{(\alpha)}_1(\sigma^{i-1} U)
\right).
\end{equation}
In particular, consider the case of $n=2$.
By the reason that will be shown in Remark \ref{rem:2022_mar9_1},
we can set $u_i^{(1)}u_i^{(2)} = 1$.
So if we define $u_i = u_i^{(1)}$, then $u_i^{(2)} = 1/u_i$.
In this case one has
$e^{(1)}_1(\sigma^{i-1} U) = \sum_{j=0}^{L-1} (u_{i+j})^{(-1)^j} = u_i + \sum_{j=1}^{L-1} (1/u_{i+j})^{(-1)^{j-1}}$
and
$e^{(1)}_1(\sigma^{i} U) = \sum_{j=0}^{L-1} (u_{i+j+1})^{(-1)^j} = \sum_{j=1}^{L} (u_{i+j})^{(-1)^{j-1}} = 
u_i + \sum_{j=1}^{L-1} (u_{i+j})^{(-1)^{j-1}}$, because $L$ is odd.
Therefore equation \eqref{eq:2022feb8_1} is written as
\begin{equation}
\frac{{\rm d} u_i}{{\rm d} t} =
u_i \sum_{j=1}^{L-1} (-1)^{j-1} \left(
u_{i+j} - \frac{1}{u_{i+j}}
\right).
\end{equation}
This is the system of differential equations that we obtained in \S 2.3.2 of reference \cite{TY21}.
\end{remark}

\subsection{Properties derived from the Lax equation}\label{subsec:2_3}
In this section, we study equation \eqref{eq:2021dec28_3} 
without assuming any explicit expression for $y^{(\alpha)}$
to find several conserved quantities.
As a byproduct, we are going to
see that this equation on the Lax matrix $\mathcal{L}$
and a matrix of the form $\mathcal{Y}$ in \eqref{eq:2021dec28_2}
basically determines the expression for $y^{(\alpha)}$ uniquely.

Equation \eqref{eq:2021dec28_3} with a matrix of the form $\mathcal{Y}$ in
\eqref{eq:2021dec28_2} is also equivalent to the set of 
differential equations
\begin{equation}\label{eq:2022feb8_2}
\frac{{\rm d} e_i^{(\alpha)}}{{\rm d} t} =  e_i^{(\alpha)}
\left(y^{(i+\alpha -L)} -
y^{(\alpha)} 
\right) +  e_{i+1}^{(\alpha)} - e_{i+1}^{(\alpha -1)}. 
\end{equation}
Letting $i=L$, we see that $e_L^{(\alpha)} (= \prod_{i=1}^L u_{i}^{(\alpha)})$
is a conserved quantity, because $e_{L+1}^{(\alpha)} =0$ for any $\alpha$.
On the other hand, if we let $i=0$ in equation \eqref{eq:2022feb8_2}, then
\begin{equation*}
0 = \frac{{\rm d} e_0^{(\alpha)}}{{\rm d} t} =  
\left(y^{(\alpha -L)} -
y^{(\alpha)} 
\right) +  e_{1}^{(\alpha)} - e_{1}^{(\alpha -1)},
\end{equation*}
because $e_0^{(\alpha)} = 1$.
Since $e_{1}^{(\alpha -1)} = e_{1}^{(\alpha -Lp)}$, this identity
is equivalent to $y^{(\alpha -L)} + e_{1}^{(\alpha)} = y^{(\alpha)}+e_{1}^{(\alpha -Lp)}$, which
determines an expression for $y^{(\alpha)}$ as 
\begin{equation}\label{eq:2022jan21_1}
y^{(\alpha)}= \sum_{j=0}^{p-1} e^{(\alpha -jL)}_1+C,
\end{equation} 
where $C$ is a term not depending on $\alpha$. 
Now by letting $i=1$ in equation \eqref{eq:2022feb8_2} we have
\begin{equation}\label{eq:2021dec28_5}
\frac{{\rm d} e_1^{(\alpha)}}{{\rm d} t} =  e_1^{(\alpha)}
\left(y^{(1+\alpha -L)} -
y^{(\alpha)} 
\right) +  e_{2}^{(\alpha)} - e_{2}^{(\alpha -1)}.
\end{equation}
Using \eqref{eq:2022jan21_1} and the condition $Lp \equiv 1 \, (\mod n)$,
one sees that
\begin{equation}\label{eq:2022jan21_2}
y^{(1+\alpha -L)} = y^{(\alpha +L(p-1))} = \sum_{j=0}^{p-1} e^{(\alpha +L(p-1-j))}_1+C =
\sum_{j=0}^{p-1} e^{(\alpha +jL)}_1+C.
\end{equation}
Then by using the identities \eqref{eq:2022jan21_1}, \eqref{eq:2021dec28_5}, \eqref{eq:2022jan21_2}, and 
$\sum_{\alpha =1}^{n} e^{(\alpha)}_1e^{(\alpha +jL)}_1=
\sum_{\alpha =1}^{n} e^{(\alpha -jL)}_1e^{(\alpha)}_1$, we have
${\rm d}(\sum_{\alpha =1}^{n}   e^{(\alpha)}_1)/{\rm d}t = 0$,
so that $\sum_{\alpha =1}^{n} e^{(\alpha)}_1$ is a conserved quantity.
Therefore, our choice of $C = - (p/n) \sum_{\alpha =1}^{n} e^{(\alpha)}_1$
for the $y^{(\alpha)}$ above equation \eqref{eq:2021dec28_2}
is so chosen as to make it a conserved quantity, and to impose the
condition $\sum_{\alpha =1}^{n}y^{(\alpha)} =0$.

\begin{remark}\label{rem:2022_mar9_1}
In the case of $p \leq \frac{n}{2}$, an expression for
$y^{(\alpha)}_{i+1} - y^{(\alpha)}_i$ was given by \eqref{eq:2021dec28_6}.
A similar expression for the case of $p > \frac{n}{2}$ can also be derived as
\begin{equation*}
y^{(\alpha)}_{i+1} - y^{(\alpha)}_i = 
\sum_{j=0}^{L(n-p)-1} \left( u_{i+j}^{(\alpha-j-1)} - 
u_{i+j}^{(\alpha-j)} \right) + u_{i}^{(\alpha-1)} - 
u_{i}^{(\alpha+1)}.
\end{equation*}
Using these expressions, 
one sees that the relation $\sum_{\alpha =1}^{n} (y^{(\alpha)}_{i+1} - y^{(\alpha)}_i)=0$
is satisfied for any $i$.
Therefore, by using equation \eqref{eq:2021dec28_7},
we have
${\rm d}(\log \prod_{\alpha =1}^{n} u^{(\alpha)}_{i})/{\rm d}t = 0$,
so that $\prod_{\alpha =1}^{n} u^{(\alpha)}_{i}$ is a conserved quantity for any $i$.
\end{remark}

\section{Type II differential equations}\label{sec:3}
\subsection{Definitions and the second result}\label{subsec:3_1}
As in \S \ref{subsec:2_1},
let $t \in \R$ be the time variable
and $u_i^{(\alpha)}$ be a set of
dependent variables 
labeled by $(\alpha, i) \in (\Z/n\Z)\times (\Z/L\Z)$, but now
the integers $L$ and $n$ are not necessarily coprime.
Suppose that the system of differential equations
\begin{equation}\label{eq:2}
\frac{{\rm d} u_i^{(\alpha)}}{{\rm d} t} = u_i^{(\alpha)}
\left(
\frac{1}{e^{(\alpha)}_{L-1}} \prod_{l=1}^{i-1}  u_l^{(\alpha)} \prod_{k=i+1}^{L}  u_k^{(\alpha -1)} -
\frac{1}{e^{(\alpha+1)}_{L-1}} \prod_{l=1}^{i-1}  u_l^{(\alpha+1)} \prod_{k=i+1}^{L}  u_k^{(\alpha)}
\right),
\end{equation}
is satisfied by them.

Comparing with the definition \eqref{eq:2022jan4_2},
one sees that part of the product in the first term of the right hand side 
of equation \eqref{eq:2} equals to
the $i$-th term of the summation in the definition of $e_{L-1}^{(\alpha)}$.
Therefore,
by using this equation and
the definitions of $e_L^{(\alpha)}$ and $e_{L-1}^{(\alpha)}$,
we have
\begin{equation*}
\frac{{\rm d} \log  e_L^{(\alpha)} }{{\rm d} t} =
\sum_{i=1}^L \frac{1}{u_i^{(\alpha)}} \frac{{\rm d} u_i^{(\alpha)}}{{\rm d} t} = 1-1=0.
\end{equation*}
Hence $e_L^{(\alpha)}$ is a conserved quantity for any $\alpha$.
Recall that the Lax matrix $\mathcal{L}$ was 
defined by equation \eqref{eq:Lmatrix}, and let
$\mathcal{Z}$ be the $n \times n$ matrix
\begin{equation}\label{eq:2021dec28_2xx}
\mathcal{Z}=
\begin{pmatrix}
0& z^{(2)}& & & \\
&0  & z^{(3)}& & \\
&  & \ddots &\ddots & \\
&&  & \ddots &z^{(n)}\\
 z^{(1)}/\lambda  & & &  & 0
\end{pmatrix},
\end{equation}
where 
\begin{equation}\label{eq:2022apr14_3}
z^{(\alpha)} =   
1/e^{(\alpha)}_{L-1}.
\end{equation}

\begin{theorem}\label{th:main2}
Suppose that the variables $u_i^{(\alpha)}$ are satisfying the system of
differential equations \eqref{eq:2}, and that the condition
$e_L^{(\alpha)} =1$ is satisfied for any $\alpha$.
Then the Lax matrix $\mathcal{L}$ satisfies the equation
\begin{equation}\label{eq:2021dec28_3xx}
\frac{{\rm d} \mathcal{L}}{{\rm d} t} = 
[\mathcal{L}, \mathcal{Z}].
\end{equation}
\end{theorem}

By Theorems \ref{th:main} and \ref{th:main2}, we see that
the dynamical systems 
represented by equations \eqref{eq:1} and 
\eqref{eq:2}
share a common Lax matrix defined by \eqref{eq:Lmatrix} in some cases.
As a result, they share a common set of conserved quantities in such cases.

\subsection{Proof of Theorem \ref{th:main2}}\label{subsec:3_2}
For the $z^{(\alpha)}$ defined in \eqref{eq:2022apr14_3}, we write its dependence on
the variable $U$ in \eqref{eq:2022jan4_1} as $z^{(\alpha)}(U)$, and define
\begin{equation}\label{eq:2022apr14_4}
z^{(\alpha)}_i = z^{(\alpha)}(\sigma^{i-1} U),
\end{equation}
for any $i \in \Z/L\Z$.
Let $\mathcal{M}_i$ be the matrix in \eqref{eq:matricesBM} and
define the matrix $\mathcal{Z}_i$ as
\begin{equation}\label{eq:matrixC}
\mathcal{Z}_i=
\begin{pmatrix}
0& z^{(2)}_i & & & \\
&0  & z^{(3)}_i & & \\
&  & \ddots &\ddots & \\
&&  & \ddots &z^{(n)}_i \\
 z^{(1)}_i /\lambda  & & &  & 0
\end{pmatrix}.
\end{equation}
\begin{proposition}\label{prop:eqMZ}
The system of differential equations \eqref{eq:2} 
with the condition $e_L^{(\alpha)} =1$ is equivalent to
\begin{equation}\label{eq:Lax_triads2}
\frac{{\rm d} \mathcal{M}_i}{{\rm d} t} = \mathcal{M}_i\mathcal{Z}_{i+1}-
\mathcal{Z}_{i}\mathcal{M}_i .
\end{equation}
\end{proposition}
This together with the relations $\mathcal{L} = \mathcal{M}_1 \cdots \mathcal{M}_L$ and
$\mathcal{Z} = \mathcal{Z}_{1}$
yields the assertion of Theorem \ref{th:main2}.

\proof
The matrix elements of equation \eqref{eq:Lax_triads2} are explicitly written as
\begin{align}
u_i^{(\alpha-1)} z^{(\alpha)}_{i+1} &= u_i^{(\alpha)} z^{(\alpha)}_{i},
\label{eq:2022mar18_1}
\\
\frac{{\rm d} u_i^{(\alpha)}}{{\rm d} t}&= z^{(\alpha)}_{i+1} - z^{(\alpha+1)}_i.
\label{eq:2021dec28_7xx}
\end{align}
Thus the assertion of the proposition is a consequence of the following two lemmas,
which are satisfied by any set of variables $u_i^{(\alpha)}$ for $(\alpha, i) \in (\Z/n\Z)\times (\Z/L\Z)$ under the condition $e_L^{(\alpha)} =1$.
\qed

\begin{lemma}\label{lem:2022jan4_5}
If $e_L^{(\alpha)}$ does not depend on $\alpha$,
then the relation \eqref{eq:2022mar18_1} or equivalently 
the following relation holds:
\begin{equation}\label{eq:2022mar10_1}
e_{L-1}^{(\alpha)}(\sigma^i U) \frac{u_i^{(\alpha)}}{u_i^{(\alpha-1)}} = 
e_{L-1}^{(\alpha)}(\sigma^{i-1} U).
\end{equation}
\end{lemma}

\proof
Suppose $e_L^{(\alpha)} =C$ for any $\alpha$.
Then
\begin{align*}
e_{L-1}^{(\alpha)}(\sigma^i U) \frac{u_i^{(\alpha)}}{u_i^{(\alpha-1)}} &= 
\sum_{l=1}^L \left(
\prod_{j=1}^{l-1}  u_{j+i}^{(\alpha)} \prod_{k=l+1}^{L}  u_{k+i}^{(\alpha -1)} \right)
\frac{u_i^{(\alpha)}}{u_i^{(\alpha-1)}} \\
&=\sum_{l=1}^{L-1} \left(
\prod_{j=1}^{l-1}  u_{j+i}^{(\alpha)} \prod_{k=l+1}^{L}  u_{k+i}^{(\alpha -1)} \right)
\frac{u_i^{(\alpha)}}{u_i^{(\alpha-1)}}  + \frac{C}{u_i^{(\alpha-1)}} \\
&=\sum_{l=1}^{L-1} \left(
\prod_{j=0}^{l-1}  u_{j+i}^{(\alpha)} \prod_{k=l+1}^{L-1}  u_{k+i}^{(\alpha -1)} \right)
+ \prod_{k=1}^{L-1}  u_{k+i}^{(\alpha -1)} \\
&=\sum_{l=2}^{L} \left(
\prod_{j=1}^{l-1}  u_{j+i-1}^{(\alpha)} \prod_{k=l+1}^{L}  u_{k+i-1}^{(\alpha -1)} \right)
+ \prod_{k=2}^{L}  u_{k+i-1}^{(\alpha -1)} \\
&=\sum_{l=1}^{L} \left(
\prod_{j=1}^{l-1}  u_{j+i-1}^{(\alpha)} \prod_{k=l+1}^{L}  u_{k+i-1}^{(\alpha -1)} \right)
=e_{L-1}^{(\alpha)}(\sigma^{i-1} U).
\end{align*}
In the second line, we wrote the term for $l=L$ separately.
In the third line, the extra factor is absorbed in the parenthesized expression,
and we used $C=e_L^{(\alpha -1)}$ in the separated term.
In the fourth line, we replaced $(j,k,l)$ by $(j-1,k-1,l-1)$.
In the last line, we included the separated term into the summation
as the term for $l=1$.
\qed

\begin{lemma}\label{lem:2022jan4_6}
Suppose $e_L^{(\alpha)} =1$ for any $\alpha$, and
let $z^{(\alpha)}_{i}$ be the one defined in \eqref{eq:2022apr14_4}.
Then
\begin{equation}
z^{(\alpha)}_{i+1} - z^{(\alpha+1)}_i =
u_i^{(\alpha)}
\left(
\frac{1}{e^{(\alpha)}_{L-1}} \prod_{l=1}^{i-1}  u_l^{(\alpha)} \prod_{k=i+1}^{L}  u_k^{(\alpha -1)} -
\frac{1}{e^{(\alpha+1)}_{L-1}} \prod_{l=1}^{i-1}  u_l^{(\alpha+1)} \prod_{k=i+1}^{L}  u_k^{(\alpha)}
\right).
\end{equation}
\end{lemma}

\proof
By definition, we have
\begin{align*}
z^{(\alpha)}_{i+1} - z^{(\alpha+1)}_i &=\frac{1}{e_{L-1}^{(\alpha)}(\sigma^i U) }-
\frac{1}{e_{L-1}^{(\alpha+1)}(\sigma^{i-1} U) }\\
&=u_i^{(\alpha)} \left(  \frac{1}{u_i^{(\alpha)}e_{L-1}^{(\alpha)}(\sigma^i U) }-
\frac{1}{u_i^{(\alpha+1)} e_{L-1}^{(\alpha+1)}(\sigma^{i} U) }\right),
\end{align*}
where we used Lemma \ref{lem:2022jan4_5} in the second line.
Therefore it suffices to show that
\begin{equation*}
u_i^{(\alpha)}e_{L-1}^{(\alpha)}(\sigma^i U) =
\left(
\prod_{l=1}^{i-1}  u_l^{(\alpha)} \prod_{k=i+1}^{L}  u_k^{(\alpha -1)}
\right)^{-1} 
e^{(\alpha)}_{L-1}(U),
\end{equation*}
for any $\alpha \in \Z/(n\Z)$ and $1 \leq i \leq L$.
This relation is proved by applying
Lemma \ref{lem:2022jan4_5} on the left hand side repeatedly as,
\begin{align*}
u_i^{(\alpha)}e_{L-1}^{(\alpha)}(\sigma^i U) &=
u_i^{(\alpha-1)}e_{L-1}^{(\alpha)}(\sigma^{i-1} U) \\
&= \frac{u_i^{(\alpha-1)} u_{i-1}^{(\alpha-1)} \dots u_1^{(\alpha-1)}}{u_{i-1}^{(\alpha)} \dots u_1^{(\alpha)}}
e_{L-1}^{(\alpha)}(U) \\
&=\left(
\prod_{l=1}^{i-1}  u_l^{(\alpha)} \prod_{k=i+1}^{L}  u_k^{(\alpha -1)}
\right)^{-1} 
e^{(\alpha)}_{L-1}(U),
\end{align*}
where we used $\prod_{k=1}^{L}  u_k^{(\alpha -1)}=e_L^{(\alpha-1)} =1$.
The proof is completed.
\qed

\begin{remark}
There is a simple expression for the system of differential equations \eqref{eq:2}
that can be compared with equation \eqref{eq:2022feb8_1} in the type I case.
In fact,
from the above considerations we see that equation \eqref{eq:2} is written as
\begin{equation}
\frac{{\rm d} u_i^{(\alpha)}}{{\rm d} t} = \frac{1}{e_{L-1}^{(\alpha)}(\sigma^i U) }-
\frac{1}{e_{L-1}^{(\alpha+1)}(\sigma^{i-1} U) }.
\end{equation}
In particular, consider the case of $L=2$.
If we define $u^{(\alpha)} = u_1^{(\alpha)}$, then we have $u_2^{(\alpha)} = 1/u^{(\alpha)}$
because we set $e_{L}^{(\alpha)} = 1$.
Therefore one has
\begin{equation}\label{eq:2022_mar4_1}
\frac{{\rm d} u^{(\alpha)}}{{\rm d} t} =
\frac{u^{(\alpha)}}{u^{(\alpha)}u^{(\alpha-1)}+1}-
\frac{u^{(\alpha)}}{u^{(\alpha)}u^{(\alpha+1)}+1} .
\end{equation}
This is basically the same differential equation that
was appeared in reference \cite{Suris97}
as equation (13.58), where its discretization was referred to as a lattice KdV equation.

\end{remark}

\subsection{Properties derived from the Lax equation}\label{subsec:3_3}
In this section, we study equation \eqref{eq:2021dec28_3xx} 
without assuming any explicit expression for $z^{(\alpha)}$
and find several conserved quantities.
As a byproduct, we are going to
see that this equation on the Lax matrix $\mathcal{L}$
and a matrix of the form $\mathcal{Z}$ in
\eqref{eq:2021dec28_2xx}
basically determines the expression for $z^{(\alpha)}$ uniquely.

In view of the definition \eqref{eq:Lmatrix}, one sees that
equation \eqref{eq:2021dec28_3xx} with a matrix of the form $\mathcal{Z}$ in
\eqref{eq:2021dec28_2xx} is equivalent to the system of 
differential equations
\begin{equation}\label{eq:2022jan5_1}
\frac{{\rm d} e_i^{(\alpha)}}{{\rm d} t} =  
z^{(i+\alpha-L)} e_{i-1}^{(\alpha)} -
z^{(\alpha+1)} e_{i-1}^{(\alpha +1)}.
\end{equation}
By letting $i=L+1$ we see that $e_L^{(\alpha)} = e_L^{(\alpha+1)}$, because $e_{L+1}^{(\alpha)} =0$ for any $\alpha$.
On the other hand, 
the summation $\sum_{\alpha=1}^ne^{(\alpha)}_{L}$ is a conserved quantity,
because the conservation of 
$\mathrm{Tr}\, \mathcal{L}=
\sum_{m \geq 0} \sum_{\alpha=1}^ne^{(\alpha)}_{L-mn}  \lambda^m$
is derived from equation \eqref{eq:2021dec28_3xx}.
This forces each $e_L^{(\alpha)}$ to be a conserved quantity.
Then by letting $i=L$ in equation \eqref{eq:2022jan5_1} we have
\begin{equation*}
0 = \frac{{\rm d} e_L^{(\alpha)}}{{\rm d} t} =  
z^{(\alpha)} e_{L-1}^{(\alpha)} -
z^{(\alpha+1)} e_{L-1}^{(\alpha +1)}.
\end{equation*}
This identity 
determines an expression for $z^{(\alpha)}$ as 
$z^{(\alpha)}= C/e_{L-1}^{(\alpha)}$, where $C$ is a factor  not depending on $\alpha$

Now letting $i=L-1$ in equation \eqref{eq:2022jan5_1} and dividing it by $e_{L-1}^{(\alpha)}$,
we have
\begin{equation*}
\frac{1}{e_{L-1}^{(\alpha)}} \frac{{\rm d} e_{L-1}^{(\alpha)}}{{\rm d} t} = C \left(  
\frac{e_{L-2}^{(\alpha)}}{e_{L-1}^{(\alpha-1)}e_{L-1}^{(\alpha)}} -
\frac{e_{L-2}^{(\alpha+1)}}{e_{L-1}^{(\alpha)}e_{L-1}^{(\alpha+1)}} \right).
\end{equation*}
This implies that
\begin{equation*}
\frac{{\rm d} }{{\rm d} t} \log (\prod_{\alpha=1}^{n}  e_{L-1}^{(\alpha)})=
\sum_{\alpha=1}^{n} \frac{1}{e_{L-1}^{(\alpha)}} \frac{{\rm d} e_{L-1}^{(\alpha)}}{{\rm d} t} =0.
\end{equation*}
Hence $\prod_{\alpha=1}^{n}  e_{L-1}^{(\alpha)}$ is a conserved quantity.
Moreover, by using equation \eqref{eq:2022mar10_1}
we have $\prod_{\alpha=1}^{n}  e_{L-1}^{(\alpha)} (\sigma^i U) =
\prod_{\alpha=1}^{n}  e_{L-1}^{(\alpha)} $ for any $i$.

\begin{remark}
We have shown the conservation of $e_L^{(\alpha)} = \prod_{i=1}^{L}  u_i^{(\alpha)}$
 for any $1 \leq \alpha \leq n$ by using equation \eqref{eq:2}.
In the same way, we can show that
\begin{equation*}
\frac{{\rm d} }{{\rm d} t} \log (\prod_{\alpha=1}^{n}  u_i^{(\alpha)})=
\sum_{\alpha=1}^{n} \frac{1}{u_i^{(\alpha)}} \frac{{\rm d} u_i^{(\alpha)}}{{\rm d} t} =0.
\end{equation*}
Hence $\prod_{\alpha=1}^{n}  u_i^{(\alpha)}$ is also a conserved quantity for any $1 \leq i \leq L$.
\end{remark}

\section{Connection to the closed geometric crystal chains: A case study for $\mathbf{ n=4}$}\label{sec:4}
\subsection{A review of the closed geometric crystal chains}\label{subsec:4_1}
We briefly review on the closed geometric crystal chains for the totally one-row tableaux case
(\cite{TY21}, \S 3.~1).
Based on the notion of {\em rational rectangles} in \cite{F19, F21}, we 
introduce the set
$\mathbb{Y}_1 = (\R_{>0})^{n-1} \times \R_{>0}$.
Let $(\mathbf{x}, s)$ denote
an element of $\mathbb{Y}_1$
with $\mathbf{x} = (x^{(1)},\dots,x^{(n-1)})$, and let $x^{(n)}:=s/(x^{(1)} \cdots x^{(n-1)})$.
Furthermore, we define $x^{(i)}$ for arbitrary $i \in \Z$ to be a variable determined
from $\mathbf{x}$ by the relation  $x^{(i)} = x^{(i+n)}$. 
In what follows, we set $n=4$.
Given $(\mathbf{x}, s) \in (\R_{>0})^{3} \times \R_{>0}$, we define 
the matrices $g^*$ and $g$ by
\begin{align}
g^*(\mathbf{x}, s; \lambda) &= 
\begin{pmatrix}
x^{(1)}  x^{(2)} x^{(3)}& \lambda &  \lambda x^{(3)}&  \lambda x^{(2)} x^{(3)}  \\
x^{(1)}  x^{(2)}  &x^{(4)}  x^{(1)}  x^{(2)} &  \lambda&   \lambda x^{(2)}\\
x^{(1)}  & x^{(4)}  x^{(1)}   & x^{(3)} x^{(4)}  x^{(1)}  & \lambda \\
1 &x^{(4)}  &  x^{(3)} x^{(4)} &x^{(2)} x^{(3)} x^{(4)} 
\end{pmatrix},\label{eq:matrix_gstar}\\
g(\mathbf{x}, s; \lambda) &=
\begin{pmatrix}
x^{(1)} &0 & 0& \lambda \\
1 & x^{(2)} &0 &0 \\
 0& 1 & x^{(3)} & 0\\
0& 0& 1 & x^{(4)} 
\end{pmatrix}.\label{eq:matrix_g}
\end{align}
Actually, any element of the matrix $g^*(\mathbf{x}, s; \lambda)$  is 
so defined as to be an order $3$ minor of the matrix $g(\mathbf{x}, s; \lambda)$. 
For instance, the top-left element of the former is equal to
the determinant of the top-left $3 \times 3$ submatrix of the latter. 

By 
Theorem 16 of \cite{TY21}, we see that
for any $s,l \in \R_{>0}$ and $(\mathbf{b}_1, \dots, \mathbf{b}_L) 
\in  (\R_{>0})^{3L} $,
there is a unique positive real solution 
$(\mathbf{v}, \mathbf{b}'_1, \dots, \mathbf{b}'_L) \in(\R_{>0})^{3(L+1)}$ to the equation
\begin{equation}\label{eq:nov20_1}
g(\mathbf{b}_1, s; \lambda) \cdots g(\mathbf{b}_L, s; \lambda) g(\mathbf{v}, l; \lambda) =
g(\mathbf{v}, l; \lambda) g(\mathbf{b}'_1, s; \lambda) \cdots g(\mathbf{b}'_L, s; \lambda).
\end{equation}
For any $| \mathbf{b} \rangle = (\mathbf{b}_1, \dots, \mathbf{b}_L)$,
let $\mathcal{L}(| \mathbf{b} \rangle ; \lambda) $ be the matrix
\begin{equation}\label{eq:sep29_2}
\mathcal{L}(| \mathbf{b} \rangle ; \lambda) =
g(\mathbf{b}_{1}, s; \lambda)
\cdots
g(\mathbf{b}_{L}, s; \lambda),
\end{equation}
which is called a Lax matrix, and let
$\mathsf{M}_l^{(1)}(| \mathbf{b} \rangle)$ be the matrix
\begin{equation}\label{eq:monodromy}
\mathsf{M}_l^{(1)}(| \mathbf{b} \rangle) = g^*(\mathbf{b}_{1}, s; l) \cdots g^*(\mathbf{b}_{L}, s; l),
\end{equation}
which we call the monodromy matrix of the Lax matrix 
$\mathcal{L}(| \mathbf{b} \rangle ; \lambda) $ with $\lambda = l$.
Note that every matrix element of $\mathsf{M}_l^{(1)}(| \mathbf{b} \rangle)$  is an
order $3$ minor of $\mathcal{L}(| \mathbf{b} \rangle ; l)$. 
Note also that, if the eigenvalues of $\mathcal{L}(| \mathbf{b} \rangle ; l)$ are given by
$\mu_1, \mu_2, \mu_3, \mu_4$, then the eigenvalues of
$\mathsf{M}_l^{(1)}(| \mathbf{b} \rangle)$ are given by
$\mu_1\mu_2\mu_3,\mu_1\mu_2\mu_4, \mu_1\mu_3\mu_4, \mu_2\mu_3\mu_4$,
where the multiplicity of the eigenvalues has been taken into account 
(Corollary 29 of \cite{TY21}).

Let $E$ be the largest eigenvalue in absolute value
of matrix $\mathsf{M}_l^{(1)}(| \mathbf{b} \rangle)$,
and $\vec{P}= (\mathcal{P}_1, \mathcal{P}_2, \mathcal{P}_{3}, 1)^t$
be an eigenvector corresponding to $E$.
By the Perron-Frobenius theorem,  $E$ is real positive,
$\vec{P}$ is uniquely determined, and 
$\mathcal{P}_1, \mathcal{P}_2, \mathcal{P}_{3}$ are all positive.
Then, the solution $\mathbf{v} \in(\R_{>0})^{3}$ of equation \eqref{eq:nov20_1} is given by
$\mathbf{v} = (\mathcal{P}_{3}, \mathcal{P}_{2}/\mathcal{P}_{3}, \mathcal{P}_{1}/\mathcal{P}_{2})$
(Proposition 15 of \cite{TY21}).
This unique solution $\mathbf{v}$ allows us to
define $T^{(1)}_l:  (\R_{>0})^{3L}  \rightarrow  (\R_{>0})^{3L}$
to be the map given by
\begin{equation}\label{eq:aug30_1}
T^{(1)}_l (\mathbf{b}_1, \dots, \mathbf{b}_L) = (\mathbf{b}'_1, \dots, \mathbf{b}'_L),
\end{equation}
which is called a time evolution.
Due to equation \eqref{eq:nov20_1},
this time evolution \eqref{eq:aug30_1} is governed by a
discrete time analogue of the Lax equation
\begin{equation}\label{eq:2022jan14_6}
\mathcal{L}(T^{(1)}_l | \mathbf{b} \rangle ; \lambda)=
g(\mathbf{v},l; \lambda)^{-1}
\mathcal{L}(| \mathbf{b} \rangle ; \lambda) g(\mathbf{v},l; \lambda).
\end{equation}
In what follows, we are going to show that this equation reduces to the continuous time
Lax equation \eqref{eq:2021dec28_3} in the limit $l \rightarrow \infty$,
or to equation \eqref{eq:2021dec28_3xx} in the limit $l \rightarrow 0$,
by making several reasonable assumptions.

\subsection{Limit for the type I Lax equation}\label{subsec:4_2}
\subsubsection{The case of \textit{p=1}.}
We assume $L \equiv 1 (\mod 4)$, and set $L = 4 \kappa + 1$.
Recall that $| \mathbf{b} \rangle = (\mathbf{b}_1, \dots, \mathbf{b}_L)$ and
$\mathbf{b}_i = (b_i^{(1)},b_i^{(2)},b_i^{(3)})$.
By the reason explained above Proposition \ref{prop:eqMY},
the Lax matrix \eqref{eq:sep29_2} can be identified with 
the matrix $\mathcal{L}$ 
defined by \eqref{eq:Lmatrix}, in which
the loop elementary symmetric functions are 
defined by \eqref{eq:lesf} but with the substitution $u_i^{(\alpha)} = b_i^{(\alpha)}$.
Using the explicit expression \eqref{eq:Lmatrix} and the condition $L = 4 \kappa + 1$,
we can obtain
the asymptotic form of the Lax matrix $\mathcal{L}(| \mathbf{b} \rangle ; l)$ under the 
limit $l \rightarrow \infty$ as
\begin{equation}\label{eq:2022jan14_1}
\mathcal{L}(| \mathbf{b} \rangle ; l)  \approx
\begin{pmatrix}
e_{1}^{(1)} l^{\kappa} &e_{2}^{(1)} l^{\kappa} & e_{3}^{(1)} l^{\kappa}& l^{\kappa+1} \\
l^{\kappa}  & e_{1}^{(2)} l^{\kappa}  &e_{2}^{(2)} l^{\kappa}  &e_{3}^{(2)} l^{\kappa}  \\
e_{3}^{(3)} l^{\kappa-1} & l^{\kappa}  & e_{1}^{(3)} l^{\kappa}  & e_{2}^{(3)} l^{\kappa} \\
e_{2}^{(4)} l^{\kappa-1} & e_{3}^{(4)} l^{\kappa-1} & l^{\kappa}  & e_{1}^{(4)} l^{\kappa}  
\end{pmatrix}.
\end{equation}
We assume that the eigenvalues $\eta_q \, (q \in \Z/4 \Z)$ of the Lax matrix $\mathcal{L}(| \mathbf{b} \rangle ; l)$ for sufficiently large $l$'s are
given by the Puiseux series expansion
\begin{equation}\label{eq:eigen_eta}
\eta_q = l^\kappa \sum_{m=-1}^\infty c_m \exp\left(\frac{\pi \sqrt{-1} mq}{2}\right) l^{-m/4},
\end{equation}
where $c_{-1} = 1$ and $c_0 = (\sum_{\alpha =1}^{4} e^{(\alpha)}_1)/4$.
This assumption is consistent with the relations
\begin{align*}
(s-l)^L = \det \mathcal{L}(| \mathbf{b} \rangle ; l) &=
\prod_{q=1}^4 \eta_q = -l^{4 \kappa + 1} + \mathcal{O}(l^{4\kappa}), \\
(\sum_{\alpha =1}^{4} e^{(\alpha)}_1) l^{\kappa} + \mathcal{O}(l^{\kappa -1})=
\mathrm{Tr}\, \mathcal{L}(| \mathbf{b} \rangle ; l) &=\sum_{q=1}^4 \eta_q =
4 l^{\kappa} c_0 + \mathcal{O}(l^{\kappa -1}),
\end{align*}
where $\mathcal{O}$ denotes Landau's symbol, and we used the identity $\det g(\mathbf{b}_i, s; l) = s-l$ for $1 \leq i \leq L$
and the asymptotic form \eqref{eq:2022jan14_1}.
For reference, we show an explicit derivation of such series expansions
for a simpler case of $n=2$ in \ref{app:B}.
Then, the asymptotic form of the largest eigenvalue of the monodromy matrix 
$\mathsf{M}_l^{(1)}(| \mathbf{b} \rangle)$ is given by
\begin{equation}\label{eq:2022jan14_4}
E = \eta_1 \eta_3 \eta_4 = l^{3 \kappa + \frac{3}{4}} + c_0 l^{3 \kappa + \frac{2}{4}} +
\cdots.
\end{equation}
In view of \eqref{eq:2022jan14_1},
we see that the asymptotic form of the monodromy matrix 
$\mathsf{M}_l^{(1)}(| \mathbf{b} \rangle)$ under the 
limit $l \rightarrow \infty$ is given by
\begin{equation}\label{eq:2022jan14_2}
\mathsf{M}_l^{(1)}(| \mathbf{b} \rangle) 
\approx
\begin{pmatrix}
\mathcal{O}(l^{3\kappa}) &l^{3\kappa+1} & e_{1}^{(3)} l^{3\kappa+1}& \mathcal{O}(l^{3\kappa+1}) \\
\mathcal{O}(l^{3\kappa}) & \mathcal{O}(l^{3\kappa})  &l^{3\kappa+1}  &e_{1}^{(2)} l^{3\kappa+1}  \\
e_{1}^{(1)} l^{3\kappa}& \mathcal{O}(l^{3\kappa})  & \mathcal{O}(l^{3\kappa})  &l^{3\kappa+1}  \\
* & * & *  & *
\end{pmatrix},
\end{equation}
where the last row is omitted because we do not need it.
Let $\mathsf{M}_{ij}$ denote the $ij$ element of $\mathsf{M}_l^{(1)}(| \mathbf{b} \rangle)$.
Then, the eigenvector 
$\vec{P}= (\mathcal{P}_1, \mathcal{P}_2, \mathcal{P}_{3}, 1)^t$
corresponding to the eigenvalue $E$ is determined by the linear equation
\begin{equation}\label{eq:2022jan14_3}
\begin{pmatrix}
\mathsf{M}_{11}-E &\mathsf{M}_{12} & \mathsf{M}_{13} \\
\mathsf{M}_{21} &\mathsf{M}_{22}-E & \mathsf{M}_{23}  \\
\mathsf{M}_{31} &\mathsf{M}_{32} & \mathsf{M}_{33}-E 
\end{pmatrix}
\begin{pmatrix}
\mathcal{P}_1 \\
\mathcal{P}_2 \\
\mathcal{P}_3
\end{pmatrix} =
-
\begin{pmatrix}
\mathsf{M}_{14} \\
\mathsf{M}_{24} \\
\mathsf{M}_{34}
\end{pmatrix}.
\end{equation}
Let $l = 1/\delta^4$.
By expressing the solution of this equation by the Cramer formula
and then substituting the asymptotic forms \eqref{eq:2022jan14_4}
and \eqref{eq:2022jan14_2} into it, we obtain the expressions
\begin{align}
\mathcal{P}_1  &= \frac{1}{\delta^3} + (e_{1}^{(1)}+e_{1}^{(2)}+e_{1}^{(3)}-3 c_0) \frac{1}{\delta^2}
+\mathcal{O}\left(\frac{1}{\delta}\right), \nonumber\\
\mathcal{P}_2  &= \frac{1}{\delta^2} + (e_{1}^{(1)}+e_{1}^{(2)}-2 c_0) \frac{1}{\delta}
+\mathcal{O}\left(1 \right), \nonumber\\
\mathcal{P}_3  &= \frac{1}{\delta} + (e_{1}^{(1)}- c_0) 
+\mathcal{O}\left(\delta \right). \label{eq:2022apr12_1}
\end{align}
We show a detailed derivation of this result in \ref{app:A1}.
Therefore, the asymptotic form of the matrix $g(\mathbf{v}, l; \lambda)$
with
$\mathbf{v} = (\mathcal{P}_{3}, \mathcal{P}_{2}/\mathcal{P}_{3}, \mathcal{P}_{1}/\mathcal{P}_{2})$
and $l = 1/\delta^4$ is expressed as
\begin{equation}\label{eq:2022jan14_5}
\delta \cdot g(\mathbf{v}, 1/\delta^4; \lambda) = \delta \cdot
\begin{pmatrix}
\mathcal{P}_{3}&  &  & \lambda \\
1 & \mathcal{P}_{2}/\mathcal{P}_{3}  && \\
& 1 & \mathcal{P}_{1}/\mathcal{P}_{2}  & \\
& & 1 & 1/(\delta^4\mathcal{P}_{1}) 
\end{pmatrix}=
\mathbb{I}_4 + \delta \cdot \mathcal{Y} + \mathcal{O}\left(\delta^2 \right),
\end{equation}
where $\mathcal{Y}$ is a matrix of the form \eqref{eq:2021dec28_2}
with $n=4, p=1$.

To summarize, we state the result of the above arguments
in the case of $n=4, p=1$ as:
\begin{theorem}\label{th:main3}
Let $\mathcal{L}( | \mathbf{b} \rangle ; \lambda)$ and $\mathcal{L}(T^{(1)}_l | \mathbf{b} \rangle ; \lambda)$ be denoted by $\mathcal{L}(t)$ and $\mathcal{L}(t + \delta)$,
respectively, where $l = 1/\delta^4$.
Assume that the eigenvalues of the matrix $\mathcal{L}(| \mathbf{b} \rangle ; l)$ for sufficiently large $l$'s are
given the expansion \eqref{eq:eigen_eta}.
Then, the discrete time Lax equation \eqref{eq:2022jan14_6}
reduces to the continuous time Lax equation \eqref{eq:2021dec28_3} in
the limit $\delta \rightarrow 0$.
\end{theorem}
\proof
By using the expression \eqref{eq:2022jan14_5},
we obtain
\begin{equation*}
\mathcal{L}(t + \delta) =
(\delta \cdot g(\mathbf{v},1/\delta^4; \lambda))^{-1}
\mathcal{L}(t)(\delta \cdot g(\mathbf{v},1/\delta^4; \lambda) )
= \mathcal{L}(t) + \delta \cdot [\mathcal{L}(t), \mathcal{Y} ]+ \mathcal{O}\left(\delta^2 \right),
\end{equation*}
from equation \eqref{eq:2022jan14_6}.
Hence the assertion of the theorem follows.
\qed


\subsubsection{The case of \textit{p=3}.}
We assume $L \equiv 3 (\mod 4)$, and set $L = 4 \kappa + 3$.
As in the previous case,
we can obtain
the asymptotic form of the Lax matrix $\mathcal{L}(| \mathbf{b} \rangle ; l)$ under the 
limit $l \rightarrow \infty$ as
\begin{equation}\label{eq:2022apr18_1}
\mathcal{L}(| \mathbf{b} \rangle ; l)  \approx
\begin{pmatrix}
e_{3}^{(1)} l^{\kappa} & l^{\kappa +1} & e_{1}^{(1)} l^{\kappa +1}& e_{2}^{(1)} l^{\kappa+1} \\
e_{2}^{(2)} l^{\kappa}  & e_{3}^{(2)} l^{\kappa}  &l^{\kappa +1}  &e_{1}^{(2)} l^{\kappa +1}  \\
e_{1}^{(3)} l^{\kappa} & e_{2}^{(3)}l^{\kappa}  & e_{3}^{(3)} l^{\kappa}  & l^{\kappa +1} \\
l^{\kappa} & e_{1}^{(4)} l^{\kappa} & e_{2}^{(4)}l^{\kappa}  & e_{3}^{(4)} l^{\kappa}  
\end{pmatrix}.
\end{equation}
We assume that the eigenvalues $\eta_q \, (q \in \Z/4 \Z)$ of the Lax matrix $\mathcal{L}(| \mathbf{b} \rangle ; l)$ for sufficiently large $l$'s are
given by the Puiseux series expansion
\begin{equation}\label{eq:eigen_etax}
\eta_q = l^{\kappa+\frac12} \sum_{m=-1}^\infty c_m \exp\left(\frac{\pi \sqrt{-1} (m-2)q}{2}\right) l^{-m/4},
\end{equation}
where $c_{-1} = 1$ and $c_2 = (\sum_{\alpha =1}^{4} e^{(\alpha)}_3)/4$.
This assumption is consistent with the relations
\begin{align*}
(s-l)^L = \det \mathcal{L}(| \mathbf{b} \rangle ; l) &=
\prod_{q=1}^4 \eta_q = -l^{4 \kappa + 3} + \mathcal{O}(l^{4\kappa +2}), \\
(\sum_{\alpha =1}^{4} e^{(\alpha)}_3) l^{\kappa} + \mathcal{O}(l^{\kappa -1})=
\mathrm{Tr}\, \mathcal{L}(| \mathbf{b} \rangle ; l) &=\sum_{q=1}^4 \eta_q =
4 l^{\kappa} c_2 + \mathcal{O}(l^{\kappa -1}),
\end{align*}
where we used the asymptotic form \eqref{eq:2022apr18_1}.
Then, the asymptotic form of the largest eigenvalue of the monodromy matrix 
$\mathsf{M}_l^{(1)}(| \mathbf{b} \rangle)$ is given by
\begin{equation}\label{eq:2022apr18_2}
E = \eta_1 \eta_3 \eta_4 = l^{3 \kappa + \frac{9}{4}} + c_0 l^{3 \kappa + \frac{8}{4}} +
\cdots.
\end{equation}
In view of \eqref{eq:2022apr18_1},
we see that the asymptotic form of the monodromy matrix 
$\mathsf{M}_l^{(1)}(| \mathbf{b} \rangle)$ under the 
limit $l \rightarrow \infty$ is given by
\begin{equation}\label{eq:2022apr18_3}
\mathsf{M}_l^{(1)}(| \mathbf{b} \rangle) 
\approx
\begin{pmatrix}
e_{1}^{(3)} l^{3\kappa+2}&\mathcal{O}(l^{3\kappa +2}) &  \mathcal{O}(l^{3\kappa+2}) &l^{3\kappa+3} \\
l^{3\kappa+2}  &e_{1}^{(2)} l^{3\kappa+2} &\mathcal{O}(l^{3\kappa +2}) & \mathcal{O}(l^{3\kappa +2})   \\
\mathcal{O}(l^{3\kappa +1}) &l^{3\kappa+2}  &e_{1}^{(1)} l^{3\kappa +2}&  \mathcal{O}(l^{3\kappa +2})   \\
\mathcal{O}(l^{3\kappa +1})  & \mathcal{O}(l^{3\kappa +1})  & l^{3\kappa+2}  & e_{1}^{(4)} l^{3\kappa +2}
\end{pmatrix}.
\end{equation}
From \eqref{eq:2022apr18_2} and \eqref{eq:2022apr18_3} one sees that
\begin{equation*}
0 = \det (\mathsf{M}_l^{(1)}(| \mathbf{b} \rangle) - E \mathbb{I}_4) =
(4 c_o - (e_{1}^{(1)}+e_{1}^{(2)}+e_{1}^{(3)}+e_{1}^{(4)})) l^{12\kappa +8+\frac34} + \mathcal{O}(l^{12\kappa +8+\frac24}),
\end{equation*}
which forces us to set $c_0 = (\sum_{\alpha =1}^4 e_{1}^{(\alpha)})/4$.
Let $l = 1/\delta^4$.
By expressing the solution of equation \eqref{eq:2022jan14_3} by the Cramer formula
and then substituting the asymptotic forms \eqref{eq:2022apr18_2}
and \eqref{eq:2022apr18_3} into it, we obtain the expressions
\begin{align}
\mathcal{P}_1  &= \frac{1}{\delta^3} + (e_{1}^{(3)}- c_0)  \frac{1}{\delta^2}
+\mathcal{O}\left(\frac{1}{\delta}\right), \nonumber\\
\mathcal{P}_2  &= \frac{1}{\delta^2} + (e_{1}^{(2)}+e_{1}^{(3)}-2 c_0) \frac{1}{\delta}
+\mathcal{O}\left(1 \right), \nonumber\\
\mathcal{P}_3  &= \frac{1}{\delta} +  (e_{1}^{(1)}+e_{1}^{(2)}+e_{1}^{(3)}-3 c_0) 
+\mathcal{O}\left(\delta \right). \label{eq:2022apr18_4}
\end{align}
From this result and equation \eqref{eq:2022mar8_1}, one sees that
the asymptotic form of the matrix $g(\mathbf{v}, l; \lambda)$
with
$\mathbf{v} = (\mathcal{P}_{3}, \mathcal{P}_{2}/\mathcal{P}_{3}, \mathcal{P}_{1}/\mathcal{P}_{2})$
and $l = 1/\delta^4$ is expressed as equation \eqref{eq:2022jan14_5},
in which $\mathcal{Y}$ is a matrix of the form \eqref{eq:2021dec28_2}
with $n=4, p=3$.
Hence we obtain the continuous time Lax equation \eqref{eq:2021dec28_3} in
the limit $\delta \rightarrow 0$, in the same way as for the case of $p=1$.

\subsection{Limit for the type II Lax equation}\label{subsec:4_3}
As in the previous subsection, we assume that $n=4$ but now let $L(\geq 2)$ be an arbitrary integer.  
We define
\begin{equation}
U_\lambda  =
\begin{pmatrix}
 & & & \lambda \\
1  &   &  &  \\
 & 1  &   &  \\
 &  & 1  &   
\end{pmatrix}.
\end{equation}
Consider the matrix \eqref{eq:sep29_2} with $s=1$ and
rewrite equation \eqref{eq:2022jan14_6} as
\begin{equation}\label{eq:2022jan14_6x}
\mathcal{L}(T^{(1)}_l | \mathbf{b} \rangle ; \lambda)=
(U_\lambda^{-1} g(\mathbf{v},l; \lambda))^{-1} \cdot (U_\lambda^{-1} 
\mathcal{L}(| \mathbf{b} \rangle ; \lambda) U_\lambda ) \cdot
 (U_\lambda^{-1} g(\mathbf{v},l; \lambda)).
\end{equation}
Accordingly, we adopt the interpretation
\begin{equation}\label{eq:2022jan14_6xxx}
\mathcal{L}(t) = U_\lambda^{-1} 
\mathcal{L}(| \mathbf{b} \rangle ; \lambda)U_\lambda, \qquad
\mathcal{L}(t + d \delta) = \mathcal{L}(T^{(1)}_l | \mathbf{b} \rangle ; \lambda),
\end{equation}
where $d$ is a parameter that will be determined later. 
Let them to be 
identified with the matrix $\mathcal{L}$ 
given by \eqref{eq:Lmatrix}, but in which the set of variables for 
the loop elementary symmetric functions 
\eqref{eq:lesf} 
should be interpreted as
\begin{equation}\label{eq:2022mar11_1}
u_i^{(\alpha -1)}(t) = b_i^{(\alpha)} \quad \mbox{and} \quad
u_i^{(\alpha)}(t + d \delta) = (b_i')^{(\alpha)}, 
\end{equation}
respectively.
Here we used the notations defined by
$T^{(1)}_l | \mathbf{b} \rangle = (\mathbf{b}'_1, \dots, \mathbf{b}'_L)$ and
$\mathbf{b}'_i = ((b_i')^{(1)},(b_i')^{(2)},(b_i')^{(3)})$.
%
The first equation of \eqref{eq:2022mar11_1} implies that
in the matrix $\mathcal{L}(| \mathbf{b} \rangle ; \lambda)$ 
we must replace $e^{(\alpha)}_m$ by $e^{(\alpha-1)}_m$ when we
change its variables from the $b$'s to the $u$'s. 
After this replacement in \eqref{eq:Lmatrix} 
and with the condition $e_{L}^{(\alpha)}=1$ for all $\alpha$,
the asymptotic form of the Lax matrix $\mathcal{L}(| \mathbf{b} \rangle ; l)$ under the 
limit $l \rightarrow 0$ is given by
\begin{equation}\label{eq:2022jan14_1x}
\mathcal{L}(| \mathbf{b} \rangle ; l)  \approx
\begin{pmatrix}
1 &e_{L-3}^{(4)} l & e_{L-2}^{(4)} l& e_{L-1}^{(4)} l \\
e_{L-1}^{(1)} & 1  &e_{L-3}^{(1)}  l  &e_{L-2}^{(1)}  l \\
e_{L-2}^{(2)} & e_{L-1}^{(2)}  & 1 & e_{L-3}^{(2)} l \\
e_{L-3}^{(3)} & e_{L-2}^{(3)} & e_{L-1}^{(3)}  & 1
\end{pmatrix}.
\end{equation}
Note that the condition $e_{L}^{(\alpha)}=1$ is preserved under the time evolution
$T^{(1)}_l$, which can be obtained from equation \eqref{eq:nov20_1} with $\lambda =0$.
We assume that the eigenvalues $\xi_q \, (q \in \Z/4\Z)$ of the Lax matrix $\mathcal{L}(| \mathbf{b} \rangle ; l)$ under the condition $s=1$ and
for sufficiently small $l$'s are
given by the Puiseux series expansion
\begin{equation}\label{eq:eigen_xi}
\xi_q = \sum_{m=0}^\infty d_m \exp\left(\frac{\pi \sqrt{-1} m q}{2}\right) l^{m/4},
\end{equation}
where $d_{0} = 1$.
This assumption is consistent with the relations
\begin{align*}
(1-l)^L = \det \mathcal{L}(| \mathbf{b} \rangle ; l) &=
\prod_{q=1}^4 \xi_q = (d_0)^4 + \mathcal{O}(l), \\
4 + \mathcal{O}(l)=
\mathrm{Tr}\, \mathcal{L}(| \mathbf{b} \rangle ; l) &=\sum_{q=1}^4 \xi_q =
4 d_0 + \mathcal{O}(l),
\end{align*}
where we used the identity $\det g(\mathbf{b}_i, 1; l) = 1-l$ for $1 \leq i \leq L$,
and
the expression \eqref{eq:2022jan14_1x}.
Then, the asymptotic form of the largest eigenvalue of the monodromy matrix 
$\mathsf{M}_l^{(1)}(| \mathbf{b} \rangle)$ is given by
\begin{equation}\label{eq:2022jan14_4x}
E = \xi_1 \xi_3 \xi_4 = 1 + d_1 l^{\frac{1}{4}} +
\cdots.
\end{equation}
From the expression \eqref{eq:2022jan14_1x},
we see that the asymptotic form of the monodromy matrix 
$\mathsf{M}_l^{(1)}(| \mathbf{b} \rangle)$ under the 
limit $l \rightarrow 0$ is given by
\begin{equation}\label{eq:2022jan14_2x}
\mathsf{M}_l^{(1)}(| \mathbf{b} \rangle)  \approx
\begin{pmatrix}
1&\mathcal{O}(l)  & \mathcal{O}(l) & e_{L-1}^{(4)} l \\
e_{L-1}^{(3)} & 1  &\mathcal{O}(l)  &\mathcal{O}(l)  \\
\mathcal{O}(1)& e_{L-1}^{(2)}  & 1 &\mathcal{O}(l)  \\
\mathcal{O}(1) & \mathcal{O}(1) & e_{L-1}^{(1)}   & 1
\end{pmatrix}.
\end{equation}
From \eqref{eq:2022jan14_4x} and \eqref{eq:2022jan14_2x} one sees that
\begin{equation*}
0 = \det (\mathsf{M}_l^{(1)}(| \mathbf{b} \rangle) - E \mathbb{I}_4) =
((d_1)^4 - e_{L-1}^{(1)}e_{L-1}^{(2)}e_{L-1}^{(3)}e_{L-1}^{(4)}) l + \mathcal{O}(l^{\frac{5}{4}}),
\end{equation*}
which forces us to set $d_1 = (\prod_{\alpha =1}^4 e_{L-1}^{(\alpha)})^{1/4} $.
Let $l = \delta^4$.
Again by using the solution of equation \eqref{eq:2022jan14_3} 
expressed by the Cramer formula
and then now substituting the asymptotic forms \eqref{eq:2022jan14_4x}
and \eqref{eq:2022jan14_2x} into it, we obtain the expressions
\begin{align}
\mathcal{P}_1  &= \frac{1}{d_1} e_{L-1}^{(4)} \delta^3 
+\mathcal{O}\left(\delta^4 \right), \nonumber\\
\mathcal{P}_2  &= \frac{1}{(d_1)^2} e_{L-1}^{(4)} e_{L-1}^{(3)} \delta^2 
+\mathcal{O}\left(\delta^3 \right), \nonumber\\
\mathcal{P}_3  &= \frac{1}{(d_1)^3} e_{L-1}^{(4)} e_{L-1}^{(3)} e_{L-1}^{(2)} \delta 
+\mathcal{O}\left(\delta^2 \right) =
\frac{d_1}{e_{L-1}^{(1)}} \delta  +\mathcal{O}\left(\delta^2 \right).\label{eq:2022apr12_2}
\end{align}
We show a detailed derivation of this result in \ref{app:A2}.
Therefore, the asymptotic form of the matrix $U_\lambda^{-1} g(\mathbf{v}, l; \lambda)$
with
$\mathbf{v} = (\mathcal{P}_{3}, \mathcal{P}_{2}/\mathcal{P}_{3}, \mathcal{P}_{1}/\mathcal{P}_{2})$
and $l = \delta^4$ is expressed as
\begin{equation}\label{eq:2022jan14_5x}
U_\lambda^{-1} g(\mathbf{v}, \delta^4; \lambda) =
\begin{pmatrix}
1&\mathcal{P}_{2}/\mathcal{P}_{3}  &  &  \\
 & 1  &\mathcal{P}_{1}/\mathcal{P}_{2} & \\
&  & 1 &\delta^4/\mathcal{P}_{1}  \\
\mathcal{P}_{3}/\lambda & &  & 1
\end{pmatrix}=
\mathbb{I}_4 + d_1 \delta \cdot \mathcal{Z} + \mathcal{O}\left(\delta^2 \right),
\end{equation}
where $\mathcal{Z}$ is a matrix of the form \eqref{eq:2021dec28_2xx}
with $n=4$.

To summarize, we state the result of the above arguments
in the case of $n=4, s=1$ as:
\begin{theorem}\label{th:main4}
Let $U_\lambda^{-1} 
\mathcal{L}(| \mathbf{b} \rangle ; \lambda)U_\lambda$ and $\mathcal{L}(T^{(1)}_l | \mathbf{b} \rangle ; \lambda)$ be denoted by $\mathcal{L}(t)$ and $\mathcal{L}(t + d \delta)$,
respectively, where $d=d_1 = (\prod_{\alpha =1}^4 e_{L-1}^{(\alpha)})^{1/4} $
and $l = \delta^4$.
Assume that
the condition $e_{L}^{(\alpha)}=1$ is satisfied for all $\alpha$, and
that the eigenvalues of the matrix $\mathcal{L}(| \mathbf{b} \rangle ; l)$ for sufficiently small $l$'s are
given the expansion \eqref{eq:eigen_xi}.
Then, the discrete time Lax equation \eqref{eq:2022jan14_6}
reduces to the continuous time Lax equation \eqref{eq:2021dec28_3xx} in
the limit $\delta \rightarrow 0$.
\end{theorem}
\proof
By using the expression \eqref{eq:2022jan14_5x},
we obtain
\begin{equation*}
\mathcal{L}(t + d \delta) =
(U_\lambda^{-1} g(\mathbf{v},\delta^4; \lambda))^{-1}
\mathcal{L}(t)(U_\lambda^{-1} g(\mathbf{v},\delta^4; \lambda) )
= \mathcal{L}(t) + d \delta \cdot [\mathcal{L}(t), \mathcal{Z} ]+ \mathcal{O}\left(\delta^2 \right),
\end{equation*}
from equation \eqref{eq:2022jan14_6x},
which was equivalent to equation \eqref{eq:2022jan14_6}.
Hence the assertion of the theorem follows.
\qed

\subsection{Limits for the type I and II differential equations}\label{subsec:4_4}
The discussions for
deriving the Lax equations in the previous two subsections
can be generalized to those for
derivations of equations \eqref{eq:1} and \eqref{eq:2}.
Consider the following matrix equation
\begin{equation}\label{eq:sep2_1}
g(\mathbf{b}, s; \lambda) g(\mathbf{a}, l; \lambda) =
g(\mathbf{a}', l; \lambda) g(\mathbf{b}', s; \lambda).
\end{equation}
For any $s, l \in \R_{>0}$ and
$(\mathbf{a},\mathbf{b}) \in (\R_{>0})^{6} $, 
there is a unique solution 
$(\mathbf{a}',\mathbf{b}') \in (\R_{>0})^{6} $ to 
this matrix equation.
Let
$R^{(s,l)}: (\R_{>0})^{6}  \rightarrow  (\R_{>0})^{6} $ be a rational map
given by $R^{(s,l)}:(\mathbf{b},\mathbf{a}) \mapsto (\mathbf{a}',\mathbf{b}')$.
This is the geometric $R$-matrix in the present case.
An explicit expression for the rational map is written as
\begin{align*}
a'^{(1)} &= a^{(1)} \frac{a^{(2)}a^{(3)}a^{(4)}+a^{(2)}a^{(3)}b^{(1)}+a^{(2)}b^{(4)}b^{(1)}+b^{(3)}b^{(4)}b^{(1)}}{a^{(1)}a^{(2)}a^{(3)}+a^{(1)}a^{(2)}b^{(4)}+a^{(1)}b^{(3)}b^{(4)}+b^{(2)}b^{(3)}b^{(4)}},\\
a'^{(2)} &= a^{(2)} \frac{a^{(3)}a^{(4)}a^{(1)}+a^{(3)}a^{(4)}b^{(2)}+a^{(3)}b^{(1)}b^{(2)}+b^{(4)}b^{(1)}b^{(2)}}{a^{(2)}a^{(3)}a^{(4)}+a^{(2)}a^{(3)}b^{(1)}+a^{(2)}b^{(4)}b^{(1)}+b^{(3)}b^{(4)}b^{(1)}},\\
a'^{(3)} &= a^{(3)} \frac{a^{(4)}a^{(1)}a^{(2)}+a^{(4)}a^{(1)}b^{(3)}+a^{(4)}b^{(2)}b^{(3)}+b^{(1)}b^{(2)}b^{(3)}}{a^{(3)}a^{(4)}a^{(1)}+a^{(3)}a^{(4)}b^{(2)}+a^{(3)}b^{(1)}b^{(2)}+b^{(4)}b^{(1)}b^{(2)}},\end{align*}
and $b'^{(i)} = a^{(i)}b^{(i)}/a'^{(i)}$  \cite{Y01}.
Let the map $R^{(s,l)}:(\mathbf{b},\mathbf{a}) \mapsto (\mathbf{a}',\mathbf{b}')$
be depicted as
\begin{equation*}
\batten{\mathbf{a}'}{\mathbf{b}}{\mathbf{b}'}{\mathbf{a}}.
\end{equation*}
Then the solution of equation \eqref{eq:nov20_1} satisfies the relations in
the diagram
\begin{equation}\label{eq:aug28_2}
\batten{\mathbf{v}}{\mathbf{b}_1}{\mathbf{b}_1'}{\mathbf{v}_2}\!\!\!
\batten{}{\mathbf{b}_2}{\mathbf{b}_2'}{\mathbf{v}_3}\!\!\!
\batten{}{}{}{\cdots\cdots}
\quad
\batten{}{}{}{\mathbf{v}_{L-1}}\,\,
\batten{}{\mathbf{b}_{L-1}}{\mathbf{b}_{L-1}'}{\mathbf{v}_{L}}\!\!\!
\batten{}{\mathbf{b}_L}{\mathbf{b}_L'}{\mathbf{v},}
\end{equation}
where $\mathbf{v}_i$'s are defined by the downward recursion relation
$R^{(s,l)}(\mathbf{b}_i,\mathbf{v}_{i+1})=
(\mathbf{v}_i,\mathbf{b}_{i}')$ with the initial condition $\mathbf{v}_{L+1} = \mathbf{v}$,
where $\mathbf{v}$ is determined by the Perron-Frobenius 
eigenvector of the monodromy matrix $\mathsf{M}_l^{(1)}(| \mathbf{b} \rangle)$ as 
described in \S \ref{subsec:4_1}.
As in \S \ref{subsec:2_2} let $\sigma$ denote the cyclic shift to the left,
so we have $\sigma | \mathbf{b} \rangle = (\mathbf{b}_2, \dots, \mathbf{b}_L,\mathbf{b}_1)$.
It is easy to see that an obvious generalization of equation \eqref{eq:2022jan14_6} is
\begin{equation}\label{eq:2022jan14_6xxxxxx}
\mathcal{L}(T^{(1)}_l (\sigma^{i-1} | \mathbf{b} \rangle) ; \lambda)=
g(\mathbf{v}_i,l; \lambda)^{-1}
\mathcal{L}( \sigma^{i-1}| \mathbf{b} \rangle ; \lambda) g(\mathbf{v}_i,l; \lambda).
\end{equation}
This equation implies 
that $\mathbf{v}_i$ can also be obtained in the same way as for $\mathbf{v}$ in
\S \ref{subsec:4_1}, by simply replacing $\mathsf{M}_l^{(1)}(| \mathbf{b} \rangle)$
by $\mathsf{M}_l^{(1)}(\sigma^{i-1} | \mathbf{b} \rangle)$.
Based on this fact, one can generalize equation \eqref{eq:2022jan14_5} as
\begin{equation}\label{eq:2022jan14_5xxx}
\delta \cdot g(\mathbf{v}_i, 1/\delta^4; \lambda) = 
\mathbb{I}_4 + \delta \cdot \mathcal{Y}_i + \mathcal{O}\left(\delta^2 \right),
\end{equation}
where $\mathcal{Y}_i$ is a matrix of the form \eqref{eq:matricesBM}
with $n=4, p=1$, and equation \eqref{eq:2022jan14_5x} as
\begin{equation}\label{eq:2022jan14_5xyzyz}
U_\lambda^{-1} g(\mathbf{v}_i, \delta^4; \lambda) =
\mathbb{I}_4 + d_1 \delta \cdot \mathcal{Z}_i + \mathcal{O}\left(\delta^2 \right),
\end{equation}
where $\mathcal{Z}_i$ is a matrix of the form \eqref{eq:matrixC}
with $n=4$.

First we consider type I case, where $l=1/\delta^4$.
Around each vertex in the diagram \eqref{eq:aug28_2} we have the relation
\begin{equation*}
g(\mathbf{b}_i',s; \lambda)=
(\delta \cdot g(\mathbf{v}_{i},l; \lambda))^{-1}
g(\mathbf{b}_i,s; \lambda) (\delta \cdot g(\mathbf{v}_{i+1},l; \lambda)).
\end{equation*}
Then by setting
\begin{equation*}
\mathcal{M}_i(t + \delta) = g(\mathbf{b}_i',s; \lambda), \qquad
\mathcal{M}_i(t ) = g(\mathbf{b}_i,s; \lambda), 
\end{equation*}
and using \eqref{eq:2022jan14_5xxx}, we can derive equation \eqref{eq:Lax_triads} in
the limit $\delta \rightarrow 0$.

Second we consider type II case, where $l=\delta^4$, $s=1$, and with the condition $e_{L}^{(\alpha)}=1$ for all $\alpha$.
As in the previous case, we have the relation
\begin{equation*}
g(\mathbf{b}_i',1; \lambda)=
(U_\lambda^{-1}  g(\mathbf{v}_{i},l; \lambda))^{-1}
(U_\lambda^{-1} g(\mathbf{b}_i,1; \lambda)U_\lambda) (U_\lambda^{-1}  g(\mathbf{v}_{i+1},l; \lambda)).
\end{equation*}
Then by setting
\begin{equation*}
\mathcal{M}_i(t + d_1 \delta) = g(\mathbf{b}_i',1; \lambda), \qquad
\mathcal{M}_i(t ) = U_\lambda^{-1} g(\mathbf{b}_i,1; \lambda)U_\lambda, 
\end{equation*}
and using \eqref{eq:2022jan14_5xyzyz}, we can derive equation \eqref{eq:Lax_triads2} in
the limit $\delta \rightarrow 0$.

Finally, we conclude that
equations \eqref{eq:1} for $p=1$ and \eqref{eq:2} are derived from 
the closed geometric crystal chains by
the above
arguments and using Propositions \ref{prop:eqMY} and \ref{prop:eqMZ}, respectively. 

\section{Concluding remarks}\label{sec:5}
\subsection{On a related discrete dynamical system}\label{subsec:5_1}
In order to avoid a potential confusion, we present a discussion to illustrate
the difference between the closed geometric crystal chain
and another discrete dynamical system in reference
\cite{KNY02}, which is also defined by using
the geometric $R$ matrix.

First we consider the closed geometric crystal chain.
Let
$R^{(s,l)}: (\R_{>0})^{2(n-1)}  \rightarrow  (\R_{>0})^{2 (n-1)} $ be a rational map
that is essentially the same one in \S \ref{subsec:4_4} but the condition $n=4$ has been
generalized to for an arbitrary positive integer $n \geq 2$.
Let $R_i^{(s,l)}$ be a map from $(\R_{>0})^{(L+1)(n-1)} =  (\R_{>0})^{n-1} \times \dots \times (\R_{>0})^{n-1}$ to itself,
which acts as the map $R^{(s,l)}$ on factors $i$ and $i+1$, and as the identity on
the other factors.
We define $\mathcal{R}^{(s,l)} = R_1^{(s,l)} \circ \cdots \circ R_L^{(s,l)}$,
which are maps from $ (\R_{>0})^{(L+1)(n-1)} $ to itself.
Then the diagram \eqref{eq:aug28_2}
for general $n$ implies that we have the relation
$\mathcal{R}^{(s,l)} (\mathbf{b}_1, \dots, \mathbf{b}_L, \mathbf{v}) = 
(\mathbf{v}, \mathbf{b}'_1, \dots, \mathbf{b}'_L)$, which enables us to
define
the time evolution $T^{(1)}_l:  (\R_{>0})^{(n-1)L}  \rightarrow  (\R_{>0})^{(n-1)L}$ and 
to obtain its
Lax representation,
in the same way as in \eqref{eq:aug30_1} and \eqref{eq:2022jan14_6}, respectively.
The above $\mathbf{v}$ is determined by the Perron-Frobenius 
eigenvector of the monodromy matrix $\mathsf{M}_l^{(1)}(| \mathbf{b} \rangle)$ as in the case of $n=4$ in \S \ref{subsec:4_1}.
The commutativity of the time evolutions $T^{(1)}_{l_1} \circ T^{(1)}_{l_2}=T^{(1)}_{l_2} \circ T^{(1)}_{l_1}$
is satisfied for any $l_1, l_2 \in \R_{>0}$, as a consequence of the fact that
the maps $R_i^{(s,l)}$ are obeying the Yang-Baxter relation
$R_{i+1}^{(s,l_2)} R_{i}^{(s,l_1)}  R_{i+1}^{(l_2,l_1)} = R_{i}^{(l_2,l_1)} R_{i+1}^{(s,l_1)} R_{i}^{(s,l_2)} $ and the involution
$R_{L+1}^{(l_2,l_1)} \circ R_{L+1}^{(l_1,l_2)} = {\rm Id}$.
The Lax equation \eqref{eq:2022jan14_6} implies that
the characteristic polynomial of the matrix $\mathcal{L}(| \mathbf{b} \rangle ; \lambda)$
in \eqref{eq:sep29_2} is invariant under the actions of the time
evolutions $T^{(1)}_l$s for any $l \in \R_{>0}$,
hence its coefficients are the conserved quantities of the closed geometric crystal chain.

Second we consider the discrete dynamical system in \cite{KNY02},
but slightly modified to be consistent with the notations used in this paper.
In this case, we choose the values of the parameter $s$ attached to each site
unequally.
So let $| s \rangle =(s_1, \dots, s_L)$ and
define $\mathcal{L}(| \mathbf{b} \rangle, | s \rangle ; \lambda) $ to be the matrix
\begin{equation}\label{eq:apr7_1}
\mathcal{L}(| \mathbf{b} \rangle, | s \rangle  ; \lambda) =
g(\mathbf{b}_{1}, s_1; \lambda)
\cdots
g(\mathbf{b}_{L}, s_L; \lambda),
\end{equation}
which is called a Lax matrix.
As in the previous paragraph,
let $R_i^{(s_i,s_{i+1})}$ be a map from $(\R_{>0})^{(L+1)(n-1)} =  (\R_{>0})^{n-1} \times \dots \times (\R_{>0})^{n-1}$ to itself,
which acts as the map $R^{(s_i,s_{i+1})}$ on factors $i$ and $i+1$, and as the identity on
the other factors.
More precisely, the map $R_i^{(s_i,s_{i+1})}$ acts on the set $(\mathbb{Y}_1)^L$.
Applied to $(| \mathbf{b} \rangle, | s \rangle) \in (\mathbb{Y}_1)^L$, it sends
$| \mathbf{b} \rangle = (\mathbf{b}_1, \dots,\mathbf{b}_i,\mathbf{b}_{i+1}, \dots, \mathbf{b}_L)$ to $ (\mathbf{b}_1, \dots,\mathbf{b}'_{i+1},\mathbf{b}'_{i}, \dots, \mathbf{b}_L)$
where $(\mathbf{b}'_{i+1},\mathbf{b}'_{i})=R^{(s_i,s_{i+1})}(\mathbf{b}_i,\mathbf{b}_{i+1})$,
and $| s \rangle =(s_1, \dots,s_i,s_{i+1},\dots, s_L)$ to $(s_1, \dots,s_{i+1},s_{i},\dots, s_L)$.
Let $Y_i \, (1 \leq i \leq L)$ be the map defined by
\begin{equation*}
Y_i = R_i^{(s_{i+1},s_{i})} \circ \cdots \circ R_{L-1}^{(s_{L},s_{i})} \circ \sigma \circ
R_{1}^{(s_{1},s_{i})} \circ \cdots \circ R_{i-1}^{(s_{i-1},s_{i})},
\end{equation*}
where $\sigma$ denotes the cyclic shift to the left.
If viewed as a map on the set $(\mathbb{Y}_1)^L$, it does not change 
$| s \rangle =(s_1, \dots, s_L)$.
So it can be simply regarded as a map on the set $(\R_{>0})^{(L+1)(n-1)}$
and the relation $Y_i | \mathbf{b} \rangle = | \mathbf{b}' \rangle$
can be described by the following diagram
\begin{equation*}
\batten{\mathbf{v}_1}{\mathbf{b}_1}{\mathbf{b}_1'}{\mathbf{v}_2}\!\!\!
\batten{}{}{}{\cdots\cdots}
\quad
\batten{}{\mathbf{b}_{i-1}}{\mathbf{b}_{i-1}'}{\mathbf{v}_{i}}\!\!\!
\battenshift{}{\mathbf{b}_i}{\mathbf{b}'_i}{\mathbf{v}_{i+1}}
\batten{}{\mathbf{b}_{i+1}}{\mathbf{b}_{i+1}'}{\cdots\cdots}
\quad
\batten{}{}{}{\mathbf{v}_{L}}\!\!\!
\batten{}{\mathbf{b}_L}{\mathbf{b}_L'}{\mathbf{v}_1.}
\end{equation*}
The commutativity of the rational maps $Y_{i} \circ Y_{j}=Y_{j} \circ Y_{i}$
is satisfied for any $1 \leq i, j \leq L$.
It is proved by using the property of
the maps $R_i^{(s,l)}$ obeying the Yang-Baxter relation,
and it also reflects the fact that
the $Y_i$s for $1 \leq i \leq L-1$ can be interpreted as generators of
the translation subgroup of the extended affine Weyl group $\widetilde{W}(A_{L-1}^{(1)})$
\cite{KNY02}.
The commutativity of this type of \emph{transfer maps} defined as a composition of
general Yang-Baxter maps was discussed in \cite{V05}, where 
its original idea was attributed to a result
in the classical paper \cite{Yang67} by C.~N.~Yang.
Following this historical background, 
we shall call the discrete dynamical system described by this set of rational maps $Y_i$
a \emph{birational Yang's system}, because
its tropical counterpart was referred to as 
a combinatorial Yang's system \cite{Kuniba11} by the same reason.

In order to see a Lax representation of this system,
we consider the map $Y_L$.
Let $Y_L | \mathbf{b} \rangle = | \tilde{\mathbf{b}} \rangle = (\tilde{\mathbf{b}}_1 , \dots, \tilde{\mathbf{b}}_L)$.
Then we have
\begin{equation}\label{eq:apr7_2}
g(\mathbf{b}_{1}, s_1; \lambda)
\cdots
g(\mathbf{b}_{L}, s_L; \lambda)=
g(\tilde{\mathbf{b}}_{L}, s_L; \lambda)
g(\tilde{\mathbf{b}}_{1}, s_1; \lambda)
\cdots
g(\tilde{\mathbf{b}}_{L-1}, s_{L-1}; \lambda).
\end{equation}
Comparing this with equation \eqref{eq:apr7_1}, we obtain the relation
\begin{equation}\label{eq:apr7_3}
\mathcal{L}(Y_L | \mathbf{b} \rangle, | s \rangle  ; \lambda) =
g(\tilde{\mathbf{b}}_{L}, s_L; \lambda)^{-1}
\mathcal{L}(| \mathbf{b} \rangle, | s \rangle  ; \lambda)
g(\tilde{\mathbf{b}}_{L}, s_L; \lambda),
\end{equation}
which can be regarded as a Lax equation.
By applying the cyclic shift operator $\sigma$ repeatedly, 
one can obtain Lax representations
for all the other time evolutions $Y_i$.
Therefore,
the characteristic polynomial of the matrix $\mathcal{L}(| \mathbf{b} \rangle, | s \rangle  ; \lambda)$
is invariant under the actions of the time
evolutions $Y_i$s for any $1 \leq i \leq L$,
because it is common to that of 
$\mathcal{L}(\sigma^i | \mathbf{b} \rangle, \sigma^i | s \rangle  ; \lambda)$
for all $i$'s.
Hence its coefficients are the conserved quantities of the birational Yang's system.

For the closed geometric crystal chain, the Lax matrix
$\mathcal{L}(| \mathbf{b} \rangle ; \lambda)$ is independent of
the parameter $l$, which is in the companion matrix $g(\mathbf{v},l; \lambda)$ 
of the Lax pair in equation \eqref{eq:2022jan14_6}.
Therefore, without changing the Lax matrix, 
one can consider the limit for taking $l \rightarrow \infty$ and 
that for $l \rightarrow 0$
in equation \eqref{eq:2022jan14_6}
to obtain the continuous time Lax equations.
As a result, we were able to obtain integrable differential equations
that share a Lax matrix
in common with a discrete dynamical system, at least for the cases
treated  in \S \ref{sec:4}.
In contrast, for the birational Yang's system, the Lax matrix $\mathcal{L}(| \mathbf{b} \rangle, | s \rangle  ; \lambda)$ depends on
the same parameter $s_L$ in common with the companion matrix 
$g(\tilde{\mathbf{b}}_{L}, s_L; \lambda)$ of the Lax pair
in equation \eqref{eq:apr7_3}.
{Therefore, one can not apply the same method to take the continuum limits.}
{However, it seems reasonable to expect that 
a generalization of the symmetric form of the Painlev\'e equations $P_{\rm IV}$ and $P_{\rm V}$ in \cite{NY98}, which is also regarded as a variation of extended
Lotka-Volterra systems, can be
obtained from this discrete dynamical system by taking a continuum limit through a similar method.}

\subsection{An outlook for generalizations}\label{subsec:5_2}
In \S \ref{sec:4}, we restricted ourselves to the case of $n=4$
for deriving the differential equations. 
It seems that the above derivations can also be 
applied to the case of general $n$, but a remark on the 
largest eigenvalue of the monodromy matrix may be in order.
In this generalization, the matrices corresponding to
\eqref{eq:matrix_gstar} and \eqref{eq:matrix_g} are connected by the relation where
elements of the matrix $g^*(\mathbf{x}, s; \lambda)$  are 
so defined as to be order $n-1$ minors of the matrix $g(\mathbf{x}, s; (-1)^n\lambda)$
(See \S 3.1.1 of reference \cite{TY21}). 
Consider the type I case with $L = n \kappa + 1$.
As in equation \eqref{eq:eigen_eta},
suppose that the eigenvalues $\eta_q \, (q \in \Z/n \Z)$ of the Lax matrix $\mathcal{L}(| \mathbf{b} \rangle ; l)$ for sufficiently large $l$'s are
given by the Puiseux series expansion
\begin{equation*}
\eta_q= l^\kappa \sum_{m=-1}^\infty c_m \exp\left(\frac{2 \pi \sqrt{-1} mq}{n}\right) l^{-m/n},
\end{equation*}
where $c_{-1} = 1$ and $c_0 = (\sum_{\alpha =1}^{n} e^{(\alpha)}_1)/n$.
This implies that the eigenvalues $\eta_q' \, (q \in \Z/n \Z)$ of the matrix $\mathcal{L}(| \mathbf{b} \rangle ; -l)$ are
given by
\begin{equation*}
\eta_q'= (-l)^\kappa \sum_{m=-1}^\infty c_m \exp\left(\frac{\pi \sqrt{-1} m(2q-1)}{n}\right) l^{-m/n}.
\end{equation*}
By the above relation between $g^*(\mathbf{x}, s; \lambda)$ and
$g(\mathbf{x}, s; (-1)^n\lambda)$, 
we see that the largest eigenvalue of the monodromy matrix
$\mathsf{M}_l^{(1)}(| \mathbf{b} \rangle)$,
which is defined similarly by equation \eqref{eq:monodromy} for general $n$, 
is given by
$E = (\prod_{q=1}^n \eta_q)/\eta_{n/2}$ for the case of even $n$, or by
$E = (\prod_{q=1}^n \eta_q')/\eta_{(n+1)/2}'$ for the case of odd $n$.
In both cases, the asymptotic form of $E$ is given by
\begin{equation*}
E =  l^{(n-1) \kappa + \frac{n-1}{n}} + c_0 l^{(n-1) \kappa + \frac{n-2}{n}} +
\cdots.
\end{equation*}
To summarize, depending on whether $n$ is even or odd,
we have to pay an attention to the choice of the sign of the loop parameter
in the matrix $\mathcal{L}(| \mathbf{b} \rangle ; (-1)^n l)$ for obtaining
the largest eigenvalue of the monodromy matrix $\mathsf{M}_l^{(1)}(| \mathbf{b} \rangle)$.

It also seems that the derivation is not restricted to
the totally one-row tableaux case but
can be generalized to the
rectangular tableaux cases in \S 3.2 of \cite{TY21}, which uses the description of
the geometric \textit{R}-matrices 
in reference \cite{F21}.
In this case, to define another time evolution $T_l^{(k)}$ we use
the monodromy matrix $\mathsf{M}_l^{(k)}(| \mathbf{b} \rangle)$
for a \textit{k}-row tableau ``carrier'', 
and the entries of this matrix
are order $n-k$ minors of the Lax matrix $\mathcal{L}(| \mathbf{b} \rangle ; (-1)^{n-k-1} l)$
with such a prescribed sign
of the loop parameter 
(\S 3.2.5 of \cite{TY21}).
Therefore, an additional attention to the choice of the sign 
of the loop parameter must be payed for
considering its largest eigenvalue $E = E_l^{(k)}$.
For instance, in the case of $k=2$, the largest eigenvalue of the monodromy matrix
$\mathsf{M}_l^{(2)}(| \mathbf{b} \rangle)$
is given by
$E = (\prod_{q=1}^n \eta'_q)/(\eta'_{n/2}\eta'_{n/2+1})$ for the case of even $n$, or by
$E = (\prod_{q=1}^n \eta_q)/(\eta_{(n-1)/2}\eta_{(n+1)/2})$ for the case of odd $n$.

Anyway, we hope that we can report a result for such cases in the near future.

\appendix

\section{Derivations of the asymptotic forms of the largest eigenvector
of the monodromy matrix}\label{app:A}
\subsection{Type I case}\label{app:A1}
Here we show a derivation for the case of $p=1$.
(The other case for $p=3$ is analogous.)
Using the asymptotic expressions
\eqref{eq:2022jan14_2} for the monodromy matrix $\mathsf{M}_l^{(1)}(| \mathbf{b} \rangle)$
and \eqref{eq:2022jan14_4} for its largest eigenvalue $E$,
we obtain
\begin{align*}
\det 
\begin{pmatrix}
\mathsf{M}_{11}-E &\mathsf{M}_{12} & \mathsf{M}_{13} \\
\mathsf{M}_{21} &\mathsf{M}_{22}-E & \mathsf{M}_{23}  \\
\mathsf{M}_{31} &\mathsf{M}_{32} & \mathsf{M}_{33}-E 
\end{pmatrix}
&\approx
(\mathsf{M}_{11}-E)(\mathsf{M}_{22}-E)(\mathsf{M}_{33}-E) 
+ \mathsf{M}_{12}\mathsf{M}_{23}\mathsf{M}_{31}\\
&= -l^{9 \kappa + \frac{9}{4}} -(3 c_0 -e^{(1)}_1) l^{9 \kappa + \frac{8}{4}} + \dots.
\end{align*}
In the same way, we have
\begin{align*}
\det 
\begin{pmatrix}
\mathsf{M}_{11}-E &\mathsf{M}_{12} & -\mathsf{M}_{14} \\
\mathsf{M}_{21} &\mathsf{M}_{22}-E & -\mathsf{M}_{24}  \\
\mathsf{M}_{31} &\mathsf{M}_{32} & -\mathsf{M}_{34}
\end{pmatrix}
&\approx
(\mathsf{M}_{11}-E)(\mathsf{M}_{22}-E)(-\mathsf{M}_{34}) \\
&= -l^{9 \kappa + \frac{10}{4}} -2 c_0 l^{9 \kappa + \frac{9}{4}} + \dots,\\
\det 
\begin{pmatrix}
\mathsf{M}_{11}-E &-\mathsf{M}_{14}  & \mathsf{M}_{13} \\
\mathsf{M}_{21} &-\mathsf{M}_{24}  & \mathsf{M}_{23}  \\
\mathsf{M}_{31} &-\mathsf{M}_{34}& \mathsf{M}_{33}-E 
\end{pmatrix}
&\approx
(\mathsf{M}_{11}-E)(-\mathsf{M}_{24})(\mathsf{M}_{33}-E) 
+ (\mathsf{M}_{11}-E)\mathsf{M}_{23}\mathsf{M}_{34}\\
&= -l^{9 \kappa + \frac{11}{4}} -(c_0 +e^{(2)}_1) l^{9 \kappa + \frac{10}{4}} + \dots,\\
\det 
\begin{pmatrix}
-\mathsf{M}_{14}  &\mathsf{M}_{12} & \mathsf{M}_{13} \\
-\mathsf{M}_{24}  &\mathsf{M}_{22}-E & \mathsf{M}_{23}  \\
-\mathsf{M}_{34}  &\mathsf{M}_{32} & \mathsf{M}_{33}-E 
\end{pmatrix}
&\approx \mathsf{M}_{12}(\mathsf{M}_{24}(\mathsf{M}_{33}-E) - \mathsf{M}_{23}\mathsf{M}_{34}) +\mathsf{M}_{13}(\mathsf{M}_{22}-E) \mathsf{M}_{34}\\
&= -l^{9 \kappa + \frac{12}{4}} -(e^{(2)}_1+e^{(3)}_1) l^{9 \kappa + \frac{11}{4}} + \dots.
\end{align*}
Therefore
\begin{align*}
\mathcal{P}_1  &= \frac{ l^{9 \kappa + \frac{12}{4}} +(e^{(2)}_1+e^{(3)}_1) l^{9 \kappa + \frac{11}{4}} + \dots}{l^{9 \kappa + \frac{9}{4}} +(3 c_0 -e^{(1)}_1) l^{9 \kappa + \frac{8}{4}} + \dots} 
= l^{\frac{3}{4}}\frac{ 1 +(e^{(2)}_1+e^{(3)}_1) l^{-\frac{1}{4}} + \dots}{1 +(3 c_0 -e^{(1)}_1) l^{- \frac{1}{4}} + \dots} \\
&=
l^{\frac{3}{4}} + (e_{1}^{(1)}+e_{1}^{(2)}+e_{1}^{(3)}-3 c_0) l^{\frac{2}{4}}
+\dots, \\
\mathcal{P}_2  &= \frac{ l^{9 \kappa + \frac{11}{4}} +(c_0 +e^{(2)}_1) l^{9 \kappa + \frac{10}{4}} + \dots}{l^{9 \kappa + \frac{9}{4}} +(3 c_0 -e^{(1)}_1) l^{9 \kappa + \frac{8}{4}} + \dots} 
= l^{\frac{2}{4}}\frac{ 1 +(c_0 +e^{(2)}_1) l^{-\frac{1}{4}} + \dots}{1 +(3 c_0 -e^{(1)}_1) l^{- \frac{1}{4}} + \dots} \\
&= l^{\frac{2}{4}} + (e_{1}^{(1)}+e_{1}^{(2)}-2 c_0) l^{\frac{1}{4}}
+\dots, \\
\mathcal{P}_3  &=\frac{ l^{9 \kappa + \frac{10}{4}} +2 c_0 l^{9 \kappa + \frac{9}{4}} + \dots}{l^{9 \kappa + \frac{9}{4}} +(3 c_0 -e^{(1)}_1) l^{9 \kappa + \frac{8}{4}} + \dots} 
= l^{\frac{1}{4}}\frac{ 1 +2 c_0 l^{-\frac{1}{4}} + \dots}{1 +(3 c_0 -e^{(1)}_1) l^{- \frac{1}{4}} + \dots} \\
&= l^{\frac{1}{4}} + (e_{1}^{(1)}- c_0) 
+\dots.
\end{align*}
Hence we obtained the result \eqref{eq:2022apr12_1} in \S \ref{subsec:4_1}.

\subsection{Type II case}\label{app:A2}
Using the asymptotic expressions
\eqref{eq:2022jan14_2x} for the monodromy matrix $\mathsf{M}_l^{(1)}(| \mathbf{b} \rangle)$
and \eqref{eq:2022jan14_4x} for its largest eigenvalue $E$,
we obtain
\begin{align*}
\det 
\begin{pmatrix}
\mathsf{M}_{11}-E &\mathsf{M}_{12} & \mathsf{M}_{13} \\
\mathsf{M}_{21} &\mathsf{M}_{22}-E & \mathsf{M}_{23}  \\
\mathsf{M}_{31} &\mathsf{M}_{32} & \mathsf{M}_{33}-E 
\end{pmatrix}
&\approx
(\mathsf{M}_{11}-E)(\mathsf{M}_{22}-E)(\mathsf{M}_{33}-E) \\
&= -d_1^3 l^{\frac{3}{4}} + \dots.
\end{align*}
In the same way, we have
\begin{align*}
\det 
\begin{pmatrix}
\mathsf{M}_{11}-E &\mathsf{M}_{12} & -\mathsf{M}_{14} \\
\mathsf{M}_{21} &\mathsf{M}_{22}-E & -\mathsf{M}_{24}  \\
\mathsf{M}_{31} &\mathsf{M}_{32} & -\mathsf{M}_{34}
\end{pmatrix}
&\approx
-\mathsf{M}_{14}\mathsf{M}_{21}\mathsf{M}_{32} 
= -e^{(4)}_{L-1} e^{(3)}_{L-1} e^{(2)}_{L-1} l  + \dots,\\
\det 
\begin{pmatrix}
\mathsf{M}_{11}-E &-\mathsf{M}_{14}  & \mathsf{M}_{13} \\
\mathsf{M}_{21} &-\mathsf{M}_{24}  & \mathsf{M}_{23}  \\
\mathsf{M}_{31} &-\mathsf{M}_{34}& \mathsf{M}_{33}-E 
\end{pmatrix}
&\approx
\mathsf{M}_{14}\mathsf{M}_{21}(\mathsf{M}_{33}-E) = -d_1  e^{(4)}_{L-1} e^{(3)}_{L-1}l^{\frac{5}{4}} + \dots,\\
\det 
\begin{pmatrix}
-\mathsf{M}_{14}  &\mathsf{M}_{12} & \mathsf{M}_{13} \\
-\mathsf{M}_{24}  &\mathsf{M}_{22}-E & \mathsf{M}_{23}  \\
-\mathsf{M}_{34}  &\mathsf{M}_{32} & \mathsf{M}_{33}-E 
\end{pmatrix}
&\approx -\mathsf{M}_{14}(\mathsf{M}_{22}-E)(\mathsf{M}_{33}-E) = -d_1^2 e^{(4)}_{L-1} l^{\frac{6}{4}} + \dots.
\end{align*}
Therefore
\begin{align*}
\mathcal{P}_1  &= \frac{ d_1^2 e^{(4)}_{L-1} l^{\frac{6}{4}} + \dots}{d_1^3 l^{\frac{3}{4}} + \dots} 
= \frac{1}{d_1} e_{L-1}^{(4)}l^{\frac{3}{4}} +\dots, \\
\mathcal{P}_2  &= \frac{ d_1  e^{(4)}_{L-1} e^{(3)}_{L-1}l^{\frac{5}{4}} + \dots}{d_1^3 l^{\frac{3}{4}} + \dots} 
= \frac{1}{(d_1)^2} e_{L-1}^{(4)} e_{L-1}^{(3)} l^{\frac{2}{4}} +\dots, \\
\mathcal{P}_3  &=\frac{ e^{(4)}_{L-1} e^{(3)}_{L-1} e^{(2)}_{L-1} l  + \dots}{d_1^3 l^{\frac{3}{4}} + \dots} 
=\frac{1}{(d_1)^3} e_{L-1}^{(4)} e_{L-1}^{(3)} e_{L-1}^{(2)}  l^{\frac{1}{4}} +\dots.
\end{align*}
Hence we obtained the result \eqref{eq:2022apr12_2} in \S \ref{subsec:4_2}.

\section{Explicit derivations of the Puiseux series expansions of the eigenvalues
of the Lax matrix for $\mathbf{ n=2}$}\label{app:B}
First we consider type I case.
In the case of $n=2$,
using the explicit expression \eqref{eq:Lmatrix} and the condition $L = 2 \kappa + 1$,
we can obtain
the asymptotic form of the Lax matrix $\mathcal{L}(| \mathbf{b} \rangle ; l)$ under the 
limit $l \rightarrow \infty$ as
\begin{equation*}
\mathcal{L}(| \mathbf{b} \rangle ; l)  =
\begin{pmatrix}
\mathcal{L}_{11} & \mathcal{L}_{12}\\
\mathcal{L}_{21}  & \mathcal{L}_{22} 
\end{pmatrix}
\approx
\begin{pmatrix}
e_{1}^{(1)} l^{\kappa} & l^{\kappa+1} \\
l^{\kappa}  & e_{1}^{(2)} l^{\kappa}  
\end{pmatrix}.
\end{equation*}
The eigenvalues of the matrix $\mathcal{L}(| \mathbf{b} \rangle ; l)$ are 
explicitly written as
\begin{equation*}
\eta_{\mp} = \frac{\mathcal{L}_{11} +\mathcal{L}_{22} \mp \sqrt{(\mathcal{L}_{11} -\mathcal{L}_{22})^2 + 4\mathcal{L}_{12}\mathcal{L}_{21} }}{2}.
\end{equation*}
Since $(\mathcal{L}_{11} -\mathcal{L}_{22})^2 + 4\mathcal{L}_{12}\mathcal{L}_{21}  = 4 l^{2 \kappa + 1} + \mathcal{O} (l^{2 \kappa})$, one has the expansion of the form
\begin{equation*}
\sqrt{(\mathcal{L}_{11} -\mathcal{L}_{22})^2 + 4\mathcal{L}_{12}\mathcal{L}_{21} } =
2 l^{\kappa + \frac12} \left(1 + \sum_{m=1}^\infty c_{2m-1} l^{-m} \right).
\end{equation*}
Also we have $\mathcal{L}_{11} +\mathcal{L}_{22} = (e_{1}^{(1)}+e_{1}^{(2)}) l^{\kappa} +  \mathcal{O} (l^{\kappa - 1})$, hence the expansion of the form
\begin{equation*}
\mathcal{L}_{11} +\mathcal{L}_{22} =
2 l^{\kappa} \left( \sum_{m=0}^\infty c_{2m} l^{-m} \right),
\end{equation*}
where $c_{2m} =0$ for sufficiently large $m$'s.
Therefore, we have the Puiseux series expansion
of the eigenvalues
$\eta_1 = \eta_-$ and $\eta_2 = \eta_+$ for sufficiently large $l$ as
\begin{equation*}
\eta_q= l^\kappa \sum_{m=-1}^\infty c_m \exp\left(\pi \sqrt{-1} mq\right) l^{-m/2},
\end{equation*}
where $c_{-1} = 1$ and $c_0 = (e_{1}^{(1)}+e_{1}^{(2)})/2$.

Second we consider type II case.
In the case of $n=2$,
using the explicit expression \eqref{eq:Lmatrix} and with the condition $e_{L}^{(\alpha)}=1$,  
we can obtain
the asymptotic form of the Lax matrix $\mathcal{L}(| \mathbf{b} \rangle ; l)$ under the 
limit $l \rightarrow 0$ as
\begin{equation*}
\mathcal{L}(| \mathbf{b} \rangle ; l)  =
\begin{pmatrix}
\mathcal{L}_{11} & \mathcal{L}_{12}\\
\mathcal{L}_{21}  & \mathcal{L}_{22} 
\end{pmatrix}
\approx
\begin{pmatrix}
1 & e_{L-1}^{(2)} l \\
e_{L-1}^{(1)}  & 1 
\end{pmatrix}.
\end{equation*}
Since $(\mathcal{L}_{11} -\mathcal{L}_{22})^2 + 4\mathcal{L}_{12}\mathcal{L}_{21}  = 4 l e_{L-1}^{(1)} e_{L-1}^{(2)} + \mathcal{O} (l^{2})$, one has the expansion of the form
\begin{equation*}
\sqrt{(\mathcal{L}_{11} -\mathcal{L}_{22})^2 + 4\mathcal{L}_{12}\mathcal{L}_{21} } =
2 l^{\frac12} \left(\sqrt{e_{L-1}^{(1)} e_{L-1}^{(2)} } + \sum_{m=1}^\infty d_{2m-1} l^{m} \right).
\end{equation*}
Also we have $\mathcal{L}_{11} +\mathcal{L}_{22} = 2 +  \mathcal{O} (l)$, hence the expansion of the form
\begin{equation*}
\mathcal{L}_{11} +\mathcal{L}_{22} =
2 \left( \sum_{m=0}^\infty d_{2m} l^{m} \right),
\end{equation*}
where $d_{2m} =0$ for sufficiently large $m$'s.
Therefore, we have the Puiseux series expansion
of the eigenvalues
$\xi_1 = \eta_-$ and $\xi_2 = \eta_+$ for sufficiently small $l$ as
\begin{equation*}
\xi_q=\sum_{m=0}^\infty d_m \exp\left(\pi \sqrt{-1} mq\right) l^{m/2},
\end{equation*}
where $d_{0} = 1$ and $d_1 = \sqrt{e_{L-1}^{(1)} e_{L-1}^{(2)} }$.

\noindent
\section*{Acknowledgements}
The author thanks Masatoshi Noumi and Yasuhiko Yamada for discussions and
drawing his attention
to the difference between the two discrete dynamical systems
discussed in \S \ref{subsec:5_1}.
He also thanks Kohei Motegi for the invitation of giving a
talk at
the online workshop \emph{Combinatorial Representation Theory
and Connections with Related Fields} at RIMS, Kyoto University in November 2021,
which motivated him to initiate this work.

\vspace{0.5cm}

\end{document}